\documentclass[showpacs,preprintnumbers,amsmath,amssymb]{revtex4}
\usepackage{amsmath,amssymb,graphics,epsfig,subfigure,hyperref}
\usepackage{color}

\begin{document}
\renewcommand{\baselinestretch}{1.3}

\title{Enhanced energy extraction via magnetic reconnection in Kerr-AdS spacetime}

\author{Bo Zhao, Chao-Hui Wang, Shao-Wen Wei \footnote{Corresponding author. E-mail: weishw@lzu.edu.cn}}

\affiliation{$^{1}$ Key Laboratory of Quantum Theory and Applications of MoE, Lanzhou University, Lanzhou 730000, China\\
	$^{2}$Lanzhou Center for Theoretical Physics, Key Laboratory of Theoretical Physics of Gansu Province, School of Physical Science and Technology, Lanzhou University, Lanzhou 730000, People's Republic of China,\\
	$^{3}$Institute of Theoretical Physics $\&$ Research Center of Gravitation,
	Lanzhou University, Lanzhou 730000, People's Republic of China}

\begin{abstract}
In this paper, we study the energy extraction from Kerr-AdS black holes following the magnetic reconnection process. The parameter space regions that satisfy the energy extraction condition, as well as the efficiency and power of the extracted energy, are analyzed. The study shows that the presence of a negative cosmological constant extends the range of dominant reconnection radial locations where the energy extraction condition is met, and enables energy extraction even from black holes with relatively low spin. Furthermore, the influence of the negative cosmological constant on energy extraction is modulated by the extent of the dominant reconnection radial region: a more negative cosmological constant enhances the extracted energy, efficiency, and power, particularly for smaller dominant reconnection radii. These results demonstrate that the energy extraction from Kerr-AdS black holes is more favorable than that from their asymptotically flat counterparts. Our results highlight the crucial role of the cosmological constant in energy extraction via magnetic reconnection.
\end{abstract}



\maketitle

\section{Introduction}
Black holes, among the most enigmatic predictions of general relativity, are ultracompact objects whose extreme spacetime curvature prevents even light from escaping beyond the event horizon. Gravitational waves \cite{R1,R2,R3} and black hole images \cite{R4,R5} have recently been observed, confirming the existence of black holes. In astrophysical phenomena, astrophysical black holes are believed to be closely related to high-energy astrophysical phenomena such as active galactic nuclei (AGN) \cite{R6,R7,R8,R9} and ultraluminous X-ray binaries \cite{R10,R11,R12}. Therefore, the study of how black holes generate large amounts of relativistic jets and  ultraluminous X-ray is a highly valued and ongoing research topic. Subsequent study has shown that the rotational energy of black holes can be regarded as one of the possible sources of the vast amount of energy released by these high-energy astrophysical phenomena.

The seminal work by Penrose first showed that rotational energy extraction from Kerr black holes could occur via the Penrose process, exploiting the existence of negative energy orbits in the ergosphere \cite{R13}. Subsequent developments by Piran and Shaham introduced the collisional Penrose process in 1975 \cite{R14}, in which two particles collide within the ergosphere to produce two new particles, one of which is captured by the black hole while the other escapes the black hole and returns to infinity. This mechanism was later extended by Banados, Silk, and West, who demonstrated that extremal Kerr black holes could serve as Planck-scale particle accelerators through the Banados-Silk-West (BSW) effect to extract energy from black holes \cite{R15}. In addtion, Blandford and Znajek proposed a mechanism, known as the Blandford-Znajek mechanism, to extract rotation energy by using the interaction between a rotating black hole and its surrounding electromagnetic field, successfully explaining the phenomenon of AGN jets  \cite{R16}. Therefore, the Blandford-Znajek mechanism is now regarded as the primary mechanism responsible for powering the relativistic jets of AGN. Later, Koide and Arai suggested that magnetic reconnection in the ergospheric plasma could provide an alternative energy extraction channel \cite{R17}, while Comisso and Asenjo performed the detailed study on extracting energy from Kerr black holes via magnetic reconnection \cite{R18}. The work of Comisso and Asenjo indicates that, under specific plasma condition, magnetic reconnection processes might surpass the Blandford-Znajek mechanism in energy extraction efficiency \cite{R18}. Moreover, the detection of magnetic fields around the supermassive black hole M$87^*$ at its center \cite{R19,R20} confirms the existence of magnetic fields in the vicinity of a real black hole. This further demonstrates the feasibility of the magnetic reconnection mechanism for energy extraction. Thus, in this paper, we explore the energy extraction based on the work of Comisso and Asenjo.

The mechanism of energy extraction from a black hole via magnetic reconnection can be described as follows. In the case of a rapidly rotating black hole embedded in a magnetic field, the frame-dragging effect induces the formation of an anti-parallel magnetic field configuration in the equatorial plane \cite{R21,R22,R23,R24}. This configuration gives rise to a current sheet, which is inherently unstable. When the size of the current sheet exceeds the critical aspect ratio \cite{R25,R26,R27}, it will be destroyed by the instability of the plasma and cause the magnetic field lines with opposite directions on both sides of the current sheet to be close to each other, cutting off and rapidly reconnecting, and then forming two new magnetic field lines \cite{R28,R29,R30,R31}. Because the two new magnetic field lines are highly curved, the magnetic tension is large. Consequently, part of the plasma accelerates under the intense magnetic tension, while another part decelerates. The accelerated particles gain more kinetic energy and escape to infinity, while the decelerated particles move towards the black hole, ultimately captured by it. To the observer at infinity, the decelerated particles have negative energy, their absorption by the black hole effectively diminishes the black hole's total energy, while the escaping jet component carries away rotational energy \cite{R17,R18,R32,R33}. As the plasma moves out of the reconnection region, the magnetic tension relaxes, and the magnetic field lines deform again under the frame-dragging effect, triggering rapid magnetic reconnection \cite{R28,R17,R18} and repeating the energy extraction process. The dominant reconnection radial location of magnetic reconnection is called the X-point, which lies within the ergosphere of a rotating black hole.

In recent years, the mechanism of energy extraction via magnetic reconnection has been significantly expanded. For instance, this mechanism has been applied to more black holes for research on black hole energy extraction, such as Lorentz breaking Kerr-Sen and Kiselev black holes \cite{R34}, the spinning braneworld black hole \cite{R32}, Kerr-MOG black holes \cite{R35}, rotating hairy black hole \cite{R36}, Konoplya-Rezzolla-Zhidenko parametrized black holes \cite{R37} and so on \cite{R38,R39,R40,R41,R42,R43,R44,R45}. Analyses have also been proposed concerning energy extraction through magnetic reconnection processes in the plunging region of black holes \cite{R46,R47,R48,R49,R50,R51,R52}. These studies demonstrate that black hole parameters such as the Lorentz breaking parameter, the tidal charge, the MOdified gravity parameter, the hairy parameters, and the deformation parameters, along with properties of the plunging region, significantly influence the efficiency of energy extraction via magnetic reconnection.  Such progress has stimulated further exploration of both gravitational theories and mechanisms for extracting energy from black holes.

On the other hand, study on AdS spacetime continues to be a topic of significant interest. AdS spacetime is a solution to Einstein's field equations with a negative cosmological constant ($\Lambda<0$), characterized by unique geometric structures and physical properties \cite{R53}. It plays a crucial role in gravitational theory and related areas of theoretical physics. The AdS/CFT correspondence, proposed by Maldacena \cite{R54}, establishes a duality between gravitational theories in AdS spacetime and conformal field theories (CFT) on its boundary. This correspondence has advanced study in black hole thermodynamics and the holographic principle. The holographic nature of AdS spacetime suggests that a gravitational theory in a higher-dimensional AdS spacetime can be equivalently described by a lower-dimensional CFT, providing a crucial tool for studying quantum gravitational effects \cite{R55,R56,R57,R58}. In AdS space, a phase transition from a stable large Schwarzschild black hole to a thermal gas space, known as the Hawking-Page transition \cite{R59,R60}. According to the AdS/CFT correspondence, this phase transition is interpreted as the confinement/deconfinement phase transition. In AdS spacetime, the cosmological constant is regarded as the thermodynamic pressure and rich phase transitions are uncovered \cite{R61,R62,R63,R64,R65}. In particular, the energy extraction via magnetic reconnection in Kerr-de Sitter ($\Lambda>0$) spacetime was investigated in Ref. \cite{R33}. It was found that as the positive cosmological constant increases, a slowly rotating Kerr-de Sitter black hole can achieve energy extraction more efficiently than a Kerr black hole. Considering that the negative cosmological constant \cite{R66} makes its spacetime structure significantly different from that of Kerr black holes and Kerr-de Sitter black holes. In this paper, we attempt to investigate the energy extraction via the magnetic reconnection in the background of a Kerr-AdS black hole.

This paper is organized as follows. In Sec. \ref{iner}, we introduce the Kerr-AdS black hole spacetime and analyze the geodesic motion of photons in its vicinity, from which several characteristic orbit radii will be derived. In Sec.~\ref{eevrm}, we calculate the hydrodynamic energy-at-infinity per unit enthalpy of the surrounding plasma for the Kerr-AdS black hole and determine the condition for energy extraction. Section~\ref{paevmr} is devoted to examine the parameter space within which energy extraction is viable. In Sec.~\ref{QQQ}, we investigate the efficiency and power of energy extraction from Kerr-AdS black holes. Finally, we summarize the key results of this study in Sec.~\ref{Conclusion}.

\section{Kerr-AdS black hole and geodesics of photons }
\label{iner}

The metric of  the Kerr-AdS black hole that we will discuss can be written in the Boyer-Lindquist coordinates ($t$, $r$, $\theta$, $\phi$) as (taking the geometrized units with $G=c=1$) \cite{R66}
\begin{equation}
	\text{d}s^2=-\frac{\Delta_r-\Delta_\theta a^2\sin^2\theta}{\rho^2}\text{d}t^2+\frac{\rho^2}{\Delta_r}\text{d}r^2+\frac{\rho^2}{\Delta_\theta}\text{d}\theta^2+\frac{\sin^2\theta}{\Sigma^2\rho^2}[(r^2+a^2)^2\Delta_\theta-\Delta_ra^2\sin^2\theta]\text{d} \phi^2+\frac{2a\sin^2\theta}{\Sigma\rho^2}[(r^2+a^2)^2\Delta_\theta-\Delta_r]\text{d}t\text{d}\phi,\label{AA}
\end{equation}
where the metric functions are given by
\begin{flalign}
	&\ \Delta_r=(r^2+a^2)(1-\frac{\Lambda r^2}{3})-2Mr,\\
	&\ \Delta_\theta=1+\frac{\Lambda}{3}a^2\cos^2\theta,\\
	&\ \rho^2=r^2+a^2\cos^2\theta,\\
	&\ \Sigma=1+\frac{\Lambda a^2}{3}.
\end{flalign}
The parameters $M$ and $a$ denote the mass and spin of the black hole, respectively. For the Kerr-AdS black, the cosmological constants $\Lambda$ takes negative values. For $\Lambda=0$, the metric \eqref{AA} reduces to the standard Kerr metric.

Since the energy extraction of the mechanism we considered occurs within the region bounded by the outer horizon and the outer infinite redshift surface, known as the ergosphere, we focus on determining the outer event horizon radius and outer infinite redshift surface radius of Kerr-AdS black holes.

\begin{figure}
	\center{\includegraphics[width=6cm]{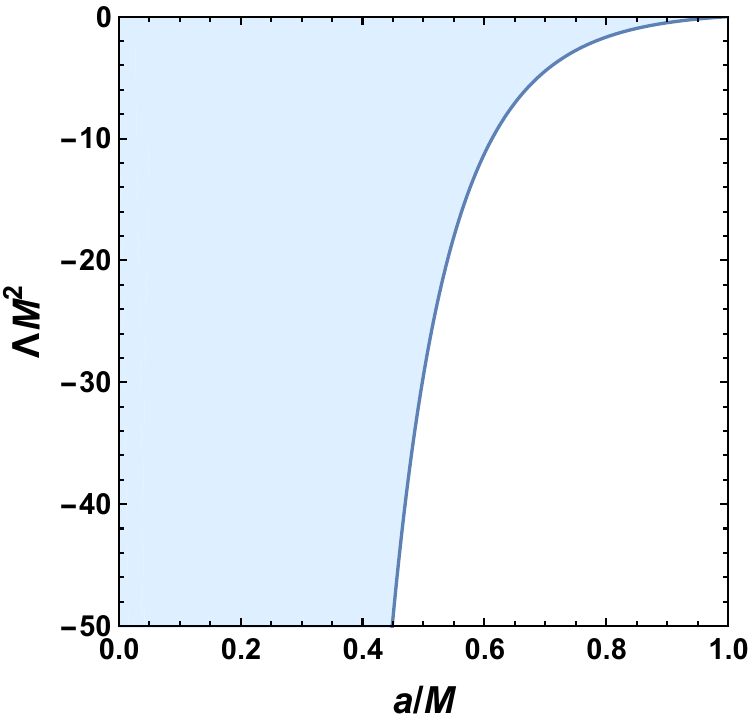}}
	\caption{Parameter space allowed the existence of Kerr-AdS black holes in $a/M-\Lambda M^2$ plane (shaded region). The solid line denotes the extremal Kerr-AdS black holes.} \label{fig1}
\end{figure}

The location of the Kerr black hole's event horizon is given by the solution of the equation $\Delta_r=0$. This is a quartic equation for $r$ with two positive roots and two negative roots. The negative roots do not have physical significance, while the two positive roots represent the inner event horizon radius $ r_-$ and the outer event horizon radius $ r_+$ of the black hole, respectively, with $ r_+ \geq r_-$. Both $ r_-$ and $ r_+$ depend on the black hole mass $M$, spin parameter $a$, and the cosmological constant $\Lambda$. Due to the complexity and length of their explicit analytical expressions, we resort to numerical methods for subsequent analysis. By solving the inequality $ r_+ \geq r_-$, we determine the parameter space of $\Lambda M^2-a/M$ that allows the existence of Kerr-AdS black holes, as illustrated in Fig. ~\ref{fig1}. In this figure, the region shaded in blue corresponds to parameter values for which Kerr-AdS black holes can exist. The solid blue curve indicates the condition $ r_+= r_-$, at which the inner and outer event horizons coincide and the black holes become extremal. Moreover, we observe that as $a/M\to 1$, $ \Lambda M^2$ approaches zero, whereas as $a/M\to 1$, $ \Lambda M^2$ can decrease without bound, tending toward negative infinity. Fig. ~\ref{fig1} shows the part of parameter space where $ \Lambda M^2$ ranges from -50 to 0.

The position of the outer boundary of the ergosphere, i.e. the outer infinite redshift surface, can be given by the following equation
\begin{equation}
	g_{tt}=-\frac{\Delta_r-\Delta_\theta a^2\sin^2\theta}{\rho^2}=0.
\end{equation}
In the paper, considering that the plasma undergoing magnetic reconnection moves in the equatorial plane, so here and after  $\theta=\frac{\pi}{2}$ is taken. It can be solved that the position of the outer ergospheres boundary $r_\text{E}$ is
\begin{equation} r_\text{E}=\frac{-3+a^2\Lambda}{(81M\Lambda^2+3\sqrt{3}\sqrt{\Lambda^3(243M^2\Lambda+(-3+a^2\Lambda)^3)})^{1/3}}-\frac{(27M\Lambda^2+\sqrt{3}\sqrt{\Lambda^3(243M^2\Lambda+(-3+a^2\Lambda)^3)})^{1/3}}{3^{2/3}\Lambda}.
\end{equation}
Since the plasma orbiting a black hole in the equatorial plane cannot reside at radii smaller than the light ring, we investigate the geodesic motion of photons in the Kerr-AdS black hole spacetime. According to the Hamilton-Jacobi equation, the geodesics of photons in this background are governed by the following expression \cite{R67,R68,R69}
\begin{flalign}
	&\  \frac{\text{d}r}{\text{d}\lambda}=\frac{\sqrt{U(r)}}{\rho^2},\\
	&\  \frac{\text{d}\theta}{\text{d}\lambda}=\frac{\sqrt{U(\theta)}}{\rho^2},\\
	&\  \frac{\text{d}\phi}{\text{d}\lambda}=\frac{a\Sigma(a^2\sin^2\theta+\rho^2)-aL\Sigma}{\rho^2\Delta_r}+\frac{L\Sigma^2-a\Sigma E\sin^2}{\rho^2\Delta_\theta sin^2},\\
	&\ \frac{\text{d}t}{\text{d}\lambda}=\frac{(a^2\sin^2\theta+\rho^2)(a^2E\sin^2\theta-aL\Sigma+E\rho^2)}{\rho^2\Delta_r}+\frac{a(L\Sigma-aE\sin^2\theta)}{\rho^2\Delta_\theta},  	
\end{flalign}
where $L$ and $E$, $\lambda$, and $K$ are the energy, angular momentum, affine parameter, and Carter constant \cite{R70}, respectively. $U(r)=[E(a^2\sin^2\theta+\rho^2)-aL\Sigma]^2-\Delta_r[(L\Sigma-aE)^2+K]$ and $U(\theta)=\Delta_\theta[(L\Sigma-aE)^2+K]-(L\Sigma csc\theta-aE\sin\theta)^2$ are non-negative definite functions of $r$ and $\theta$, respectively. A photon along the circular orbit in the equatorial plane should satisfy the following conditions
\begin{equation}
 U(r_{\text{LR}})=0,\qquad\frac{\text{d} U(r_{LR})}{\text{d} r}=0.
\end{equation}
Solving the above equation could yield the corotating orbital radius of the photon around the black hole, which we demonstrate to depend functionally on the mass $M$, spin $a$, and cosmological constants $\Lambda$. Due to the analytical expression is more complex, we employ numerical methods to investigate the orbit characteristics.

\begin{figure}
	\centering
	\subfigure[\;$a/M=0.99$]{\label{2a}
		\includegraphics[width=6cm]{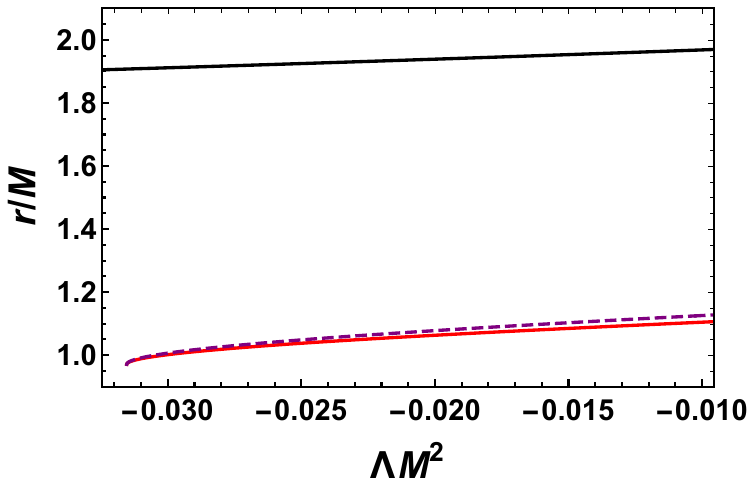}}
	\subfigure[\;$\Lambda M^2=-0.01$]{\label{2b}
		\includegraphics[width=5.8cm]{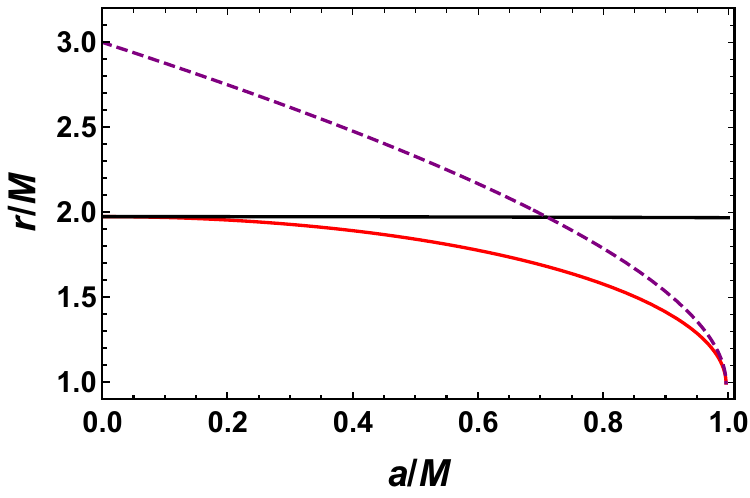}}
	\caption{The characteristic radii for the black holes. The outer event horizon $ r_+$, outer ergosphere boundary $ r_\text{E}$, and the circular corotating photon orbit $ r_{\text{LR}}$, are described by the red solid curves, black solid curves, and purple dashed curves, respectively. (a) $r/M$ vs. $\Lambda M^2$. (b) $r/M$ vs. $a/M$.} \label{fig2}
\end{figure}

We plot the radii of the outer event horizon $ r_+$ (red solid curves), the outer ergosphere boundary $ r_\text{E}$ (black solid curves), and the circular corotating photon orbit $ r_{\text{LR}}$ (purple dashed curves) as a function of the cosmological constant $ \Lambda M^2$ for a fixed spin parameter $a/M=0.99$ in Fig. \ref{2a}, and as functions of the spin parameter $a/M$ for a fixed cosmological constant $ \Lambda M^2=-0.01$ in Fig. \ref{2b}, respectively. In Fig. \ref{2a}, it shows that for a fixed $a/M$, all these three radii increase monotonically with $ \Lambda M^2$. Significantly, the circular corotating photon orbit is located very close to the outer event horizon under these conditions. Moreover, the outer ergosphere boundary consistently extends beyond both the outer horizon and the circular photon orbit. When $ \Lambda M^2$ reaches its minimum value, the black hole approaches the extremal limit, and the circular photon orbit coincides precisely with the outer event horizon. Fig.~\ref{2b} demonstrates that, for a fixed $\Lambda M^2$, all these three radii decrease as the spin parameter $a/M$ increases. When $a/M$ is small, the outer ergosphere boundary lies very close to the event horizon, and the black hole begins to resemble the Schwarzschild solution, with the ergosphere becoming increasingly narrow. As $a/M$ approaches its maximum value, the black hole becomes extremal, and again, the circular corotating photon orbit coincides with the outer horizon.

\section{Energy extraction via magnetic reconnection}
\label{eevrm}

In this section, we will calculate the hydrodynamic energy-at-infinity per enthalpy of the plasma $\epsilon^\infty_{\pm}$ surrounding the Kerr-AdS black hole and analyze the condition for energy extraction.

In order to analyze the plasma energy density conveniently, we adopt the zero-angular-momentum-observer (ZAMO) frame \cite{R71,R72}, where the line element of spacetime is written in the following form
\begin{equation}
	\text{d}s^2=-\eta_{\mu\nu}\text{d}\hat{x}^\mu \text{d}\hat{x}^\nu=-\text{d}\hat{t}^2+\Sigma^3_{i=1}(\text{d}\hat{x}^i)^2.
\end{equation}
The coordinate transformation between the Boyer-Lindquist frame and the ZAMO frame is
\begin{equation}
	\text{d}\hat{t}=\alpha \text{d}t,\quad \text{d}\hat{x}^i=\sqrt{g_{ii}}\text{d}x^i-\alpha\beta^i \text{d}t,
\end{equation}
where the lapse function $\alpha$ and the shift vector $\beta^i=(0,0,\beta^\phi)$ can be expressed as
\begin{equation}
	\alpha=\sqrt{-g_{tt}+\frac{g^2_{t\phi}}{g_{\phi\phi}}},\quad \beta^\phi=\sqrt{g_{\phi\phi}}\frac{\omega^\phi}{\alpha},
\end{equation}
with $\omega^\phi=-g_{t\phi}/g_{\phi\phi}$ being the angular velocity of the frame dragging.

For the contravariant components $a^\mu$ and the covariant components $a_\mu$ of a vector in the Boyer-Lindquist frame, the corresponding relationships when transforming to the ZAMO frame are as follows
\begin{flalign}
	&\ \hat{a}^0=\alpha a^0,\qquad\hat{a}^i=\sqrt{g_{ii}}a^i-\alpha\beta^ia^0,\\
	&\ \hat{a}_0=a_0/\alpha+\Sigma^3_{i=1}(\beta^i/g_{ii})a_i,\qquad\hat{a}_i=a_i/g_{ii}.
\end{flalign}
Next, let us analyze the energy density of the plasma. In the one-fluid approximation, the energy-momentum tensor for the plasma can be written as \cite{R17,R18,R73,R74}
\begin{equation}
	T^{\mu\nu}=pg_{\mu\nu}+\omega U^\mu U^\nu+F^\mu_\sigma F_{\nu\sigma}-\frac{1}{4}g_{\mu\nu}F^{\rho\sigma} F_{\rho\sigma},
\end{equation}
where $p$, $\omega$, $U^\mu$, and $F^{\mu\nu}$ are the proper plasma pressure, enthalpy density, four-velocity, and electromagnetic field tensor, respectively. The energy-at-infinity density is defined as follows \cite{R17,R18}
\begin{equation}
	e^\infty=-\alpha g_{\mu 0}T^{\mu 0}=\alpha \hat{e}+\alpha\beta^\phi \hat{P}^\phi.
\end{equation}
The total energy density $\hat{e}$ and the azimuthal component of the momentum density $\hat{P}^\phi$ are
\begin{flalign}
	&\ \hat{e}=\omega\hat{\gamma}^2-p+\frac{1}{2}(\hat{B}^2+\hat{E}^2),  \notag\\
	&\ \hat{P}^\phi=\omega\hat{\gamma}^2v^\phi+(\hat{B}\times\hat{E})^\phi,
\end{flalign}
where $\hat{\gamma}=\hat{U}^0=\sqrt{1-\Sigma^3_{i=1}(d\hat{v}^i)^2}$, $\hat{B}^i=\epsilon^{ijk}F_{jk}/2$, and $ \hat{E}^i=\eta^{ij}\hat{F}_{j0}=\hat{F}_{i0}\hat{v}^\phi$ \cite{R18,R75,R76} represent the Lorentz factor, the components of magnetic and the electric fields respectively, and $\hat{v}^\phi$ is the outflow velocity component of the plasma after magnetic reconnection in the ZAMO frame.

The energy-at-infinity density $e^\infty$ can be divided into two parts, the hydrodynamic component $e^\infty_{\text{hyd}}$ and electromagnetic component $e^\infty_{\text{em}}$, thus we can obtain  $e^\infty=e^\infty_{\text{hyd}}+e^\infty_{\text{em}}$ with \cite{R18}
\begin{flalign}
	&\	e^\infty_{\text{hyd}}=\alpha(\omega\hat{\gamma}^2-p) +\alpha\beta^\phi\omega\hat{\gamma}^2\hat{v}^\phi, \notag\\
	&\ e^\infty_{\text{em}}=\alpha(\hat{B}^2+\hat{E}^2)/2+\alpha\beta^\phi(\hat{B}\times\hat{E})_\phi.
\end{flalign}
It is considered that magnetic reconnection is a very efficient process, which converts most of the magnetic energy into the kinetic energy of the plasma, so that the  electromagnetic energy-at-infinity density will be insignificant. Therefore, considering  the condition that the plasma is incompressible and adiabatic, the energy-at-infinity density of the plasma  can be written as
\begin{equation}
	e^\infty=e^\infty_{\text{hyd}}=\alpha((\hat{\gamma}+\beta^\phi\hat{\gamma}\hat{v}^\phi)\omega-\frac{p}{\hat{\gamma}}).
\end{equation}
To study  the localized magnetic reconnection process, the local rest frame $x^{\prime\mu}=(x^{\prime0},x^{\prime1},x^{\prime2},x^{\prime3})$ is employed in which $ x^{\prime1}$ is parallel to the direction of $ x^1=r$ and $ x^{\prime3}$ is parallel to the direction of $ x^3=\phi$.  In the Boyer-Lindquist coordinate system, this local rest frame moves with the plasma corotating around the Kerr-AdS black hole in the circular orbit within the equatorial plane with the Kepler angular velocity $\Omega_\text{K}$, which can be given by the following equation
\begin{equation}
	\Omega_\text{K}=\frac{\text{d}\phi}{\text{d}t} =\frac{-g_{t\phi,r}+\sqrt{g^2_{t\phi,r}-g_{tt,r}{g_{\phi\phi,r}} }}{g_{\phi\phi,r}}=\frac{\sqrt{M-\frac{1}{3}r^3\Lambda } }{a\sqrt{M-\frac{1}{3}r^3\Lambda}+r^{3/2}}.
\end{equation}
In the ZAMO frame, the Keplerian angular velocity can be expressed as
\begin{equation}
	\hat{v}_\text{K}=\frac{\text{d} \hat{\phi}}{\text{d} \hat{t}}=\frac{\text{d}\hat{\phi}/\text{d}\lambda}{\text{d}\hat{t}/\text{d}\lambda}=\frac{g_{\phi\phi}\text{d}x^\phi/\text{d}\lambda-\alpha\beta^\phi \text{d}t/\text{d}\lambda}{\alpha \text{d}t/\text{d}\lambda}=\frac{\Omega_\text{K}\sqrt{g_{\phi\phi}}}{\alpha}-\beta^\phi.
\end{equation}
Then, the plasma outflow velocity in the local rest frame is expressed by $v_{\text{out}}^\prime$, which is related to the  plasma magnetization and can be expressed as \cite{R18}
\begin{equation}
	v_{\text{out}}^\prime=\sqrt{\frac{\sigma_0}{1+\sigma_0}}.
\end{equation}
\begin{figure}[htb]
	\centering
		\subfigure[\;$r/M=1.1, \xi=\pi/12$]{\label{3a}
			\includegraphics[width=4.8cm]{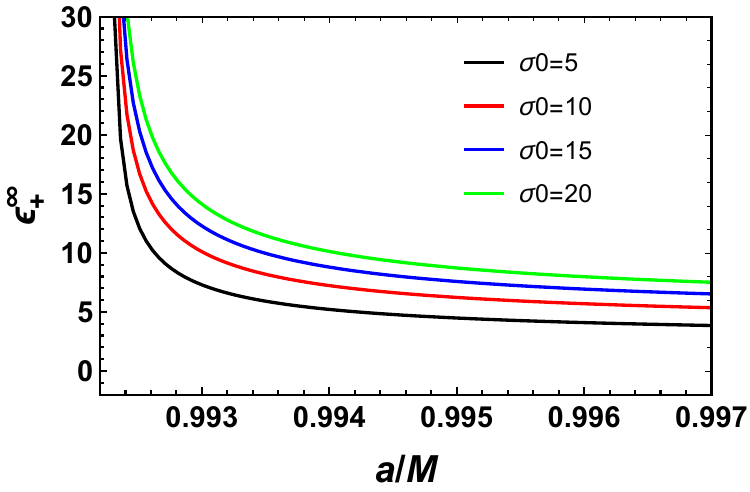}}
		\subfigure[\;$r/M=1.1, \xi=\pi/12$]{\label{3b}
			\includegraphics[width=4.8cm]{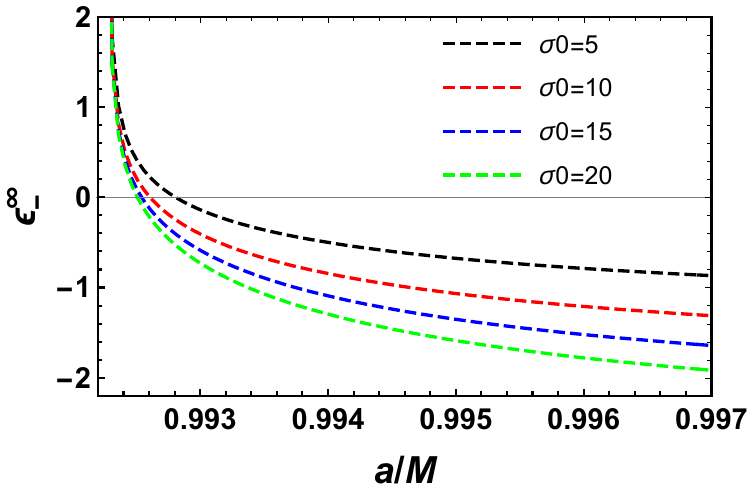}} \\
		\subfigure[\;$r/M=1.4, \xi=\pi/12$]{\label{3c}
			\includegraphics[width=4.8cm]{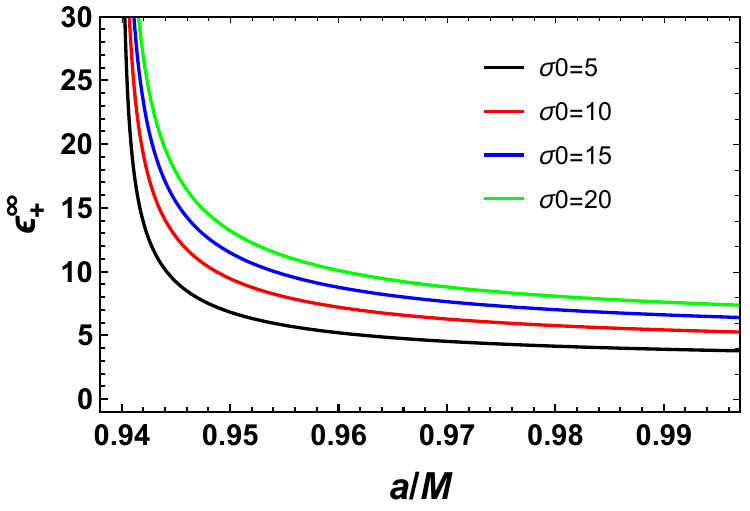}}
		\subfigure[\;$r/M=1.4, \xi=\pi/12$]{\label{3d}
			\includegraphics[width=4.8cm]{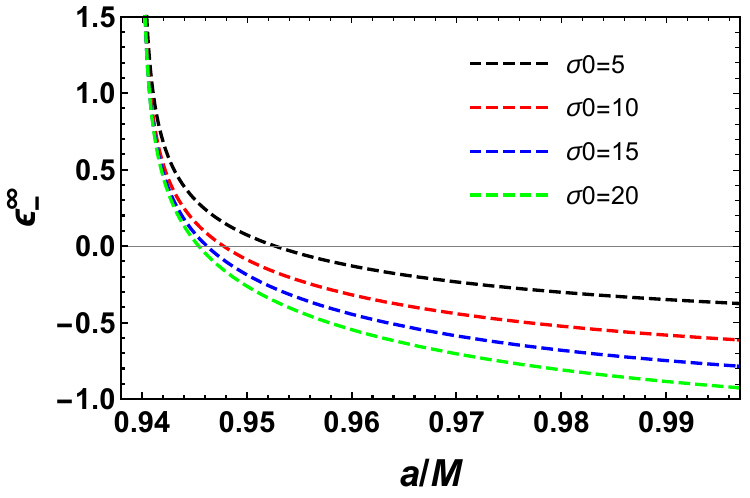}} \\
		\subfigure[\;$r/M=1.6, \xi=\pi/12$]{\label{3e}
			\includegraphics[width=4.8cm]{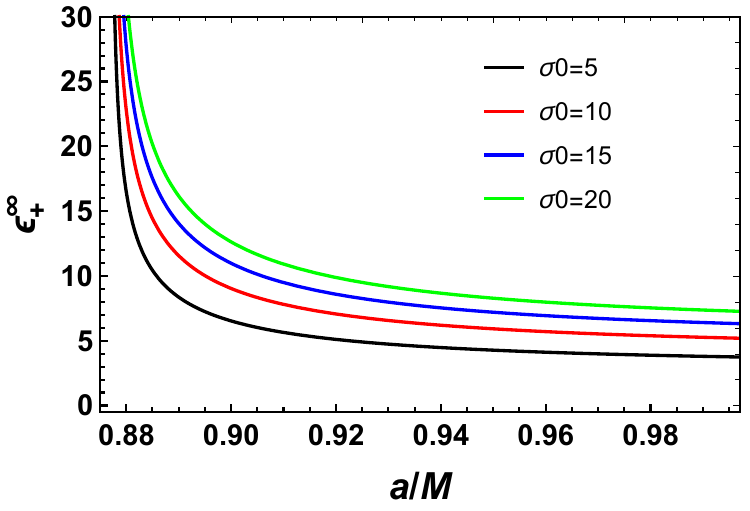}}
		\subfigure[\;$r/M=1.6, \xi=\pi/12$]{\label{3f}
			\includegraphics[width=4.8cm]{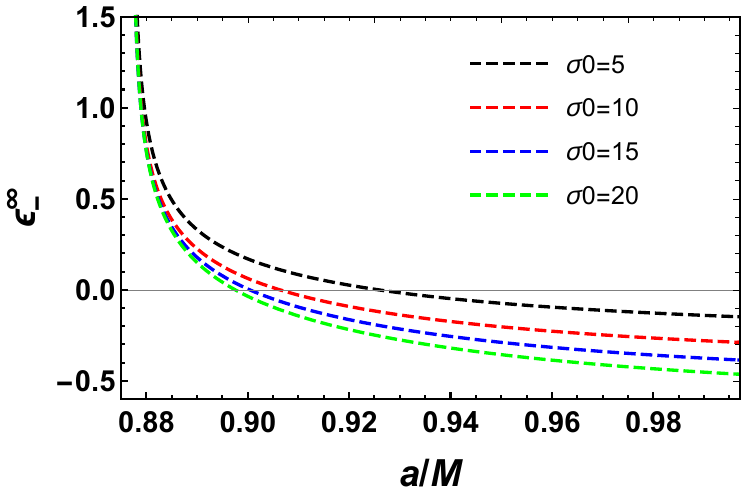}} \\
		\subfigure[\;$r/M=1.1, \xi=\pi/6$]{\label{3g}
			\includegraphics[width=4.8cm]{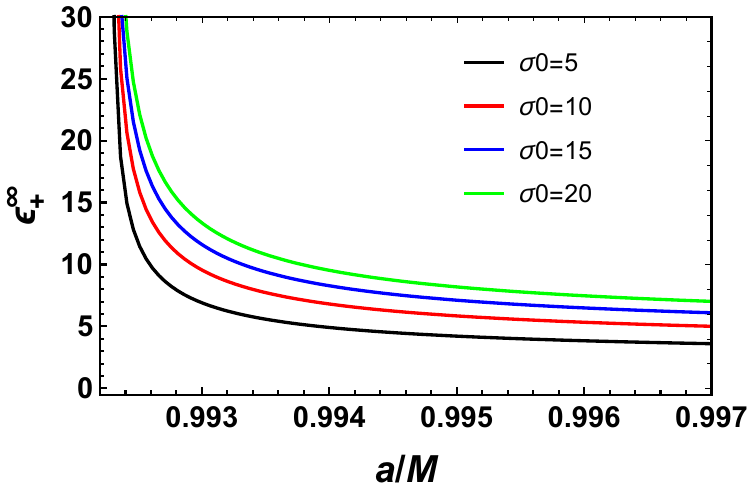}}
		\subfigure[\;$r/M=1.1, \xi=\pi/6$]{\label{3h}
			\includegraphics[width=4.8cm]{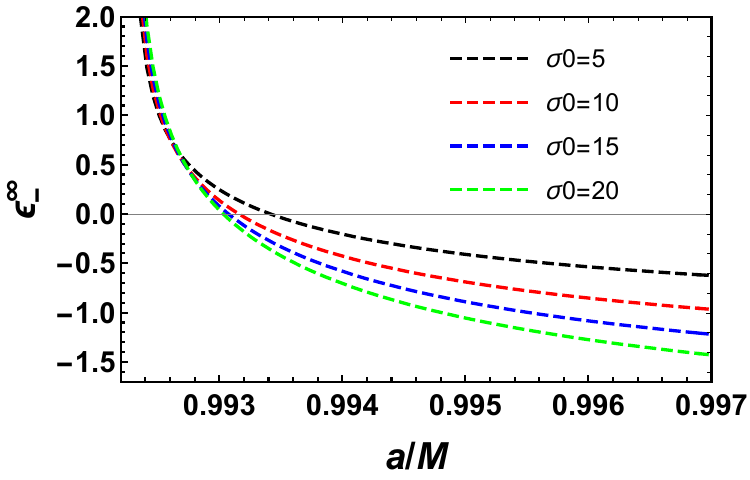}}
	\caption{The behaviors of $\epsilon^\infty_+$ (solid curves) and $\epsilon^\infty_-$ (dashed curves) as functions of spin $a/M$ for different values of the plasma magnetization of plasma $ \sigma_0$, the orientation angle $ \xi$, and the dominant reconnection radial location $r/M$, and with cosmological constant $\Lambda M^2=-0.01$ fixed. The plasma magnetization  $\sigma_0$ = 5, 10, 15, and 20 from bottom to top for solid curves and from top to bottom for dashed curves. }\label{fig3}
\end{figure}

Here $\sigma_0=B_0/2$ is the plasma magnetization upstream of the reconnection layer and $ B_0$ is the asymptotic macro-scale magnetic field. The azimuthal component of the plasma outflow velocity can be written as following formula in the ZAMO frame \cite{R18}
\begin{equation}
	\hat{v}^\phi_{\pm}=\frac{\hat{v}_\text{K}\pm v_{\text{out}}^\prime \cos\xi}{1\pm \hat{v}_\text{K}v_{\text{out}}^\prime \cos\xi},
\end{equation}
where ``+" and ``-" represent the accelerating and decelerating parts of the plasma, respectively. In addition, $ \xi=\arctan(v_{\text{out}}^{\prime 1}/v_{\text{out}}^{\prime 3})$ is  the plasma orientation angle. $ v_{\text{out}}^{\prime 1}$ and $ v_{\text{out}}^{\prime 3}$ denote the radial and azimuthal components of plasma velocities in the local rest frame, respectively.

Then, by introducing the Lorentz factor $ \hat{\gamma}_\text{K}=1/\sqrt{1-\hat{v}^2_\text{K}}$, the plasma energy-at-infinity density per enthalpy is \cite{R18}
\begin{flalign}
	\epsilon^\infty_{\pm} &=e^\infty_{\text{hyd}\pm}/\omega \notag \\
	&=\alpha \hat{\gamma}_\text{K}((1+\beta^\phi\hat{v}_\text{K})\sqrt{1+\sigma_0}\pm \cos\xi(\beta^\phi+\hat{v}_\text{K})-\frac{\sqrt{1+\sigma_0}\mp \cos\xi\hat{v}_\text{K}\sqrt{\sigma_0}}{4\hat{\gamma}_\text{K}(1+\sigma_0-\cos^2\xi\hat{v}_\text{K}^2\sigma_0)}) \label{en},
\end{flalign}
where we have assumed that the plasma is relativistic hot, so $p = \omega/4$ \cite{R18} is used. In order to achieve energy extraction, the requirements for the energy density $\epsilon^\infty_\pm$ of the decelerating particle and the accelerating particle are
\begin{equation}
	\epsilon^\infty_-<0,\qquad \Delta \epsilon^\infty_+=\epsilon^\infty_+-(1-\frac{\Gamma}{\Gamma-1}\frac{p}{\omega})=\epsilon^\infty_+>0 \label{condition},
\end{equation}
for relativistically hot plasma with polytropic index $ \Gamma=4/3$ \cite{R18}. If the decelerating plasma acquires negative energy measured at infinity, the accelerating plasma shall gain kinetic energy greater than its original energy at infinity. As a result, the rotational energy of the black hole is extracted.

\begin{figure}[htb]
	\centering
		\subfigure[\;$r/M=1.15, \xi=\pi/12$]{\label{4a}
			\includegraphics[width=4.8cm]{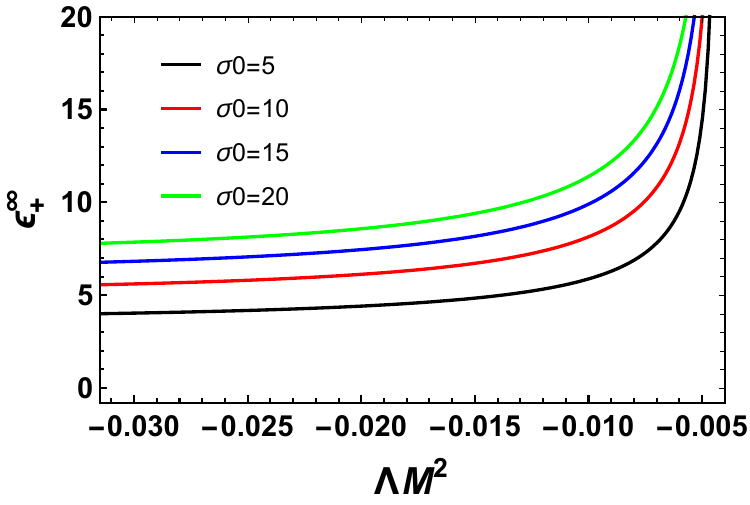}}
		\subfigure[\;$r/M=1.15, \xi=\pi/12$]{\label{4b}
			\includegraphics[width=4.8cm]{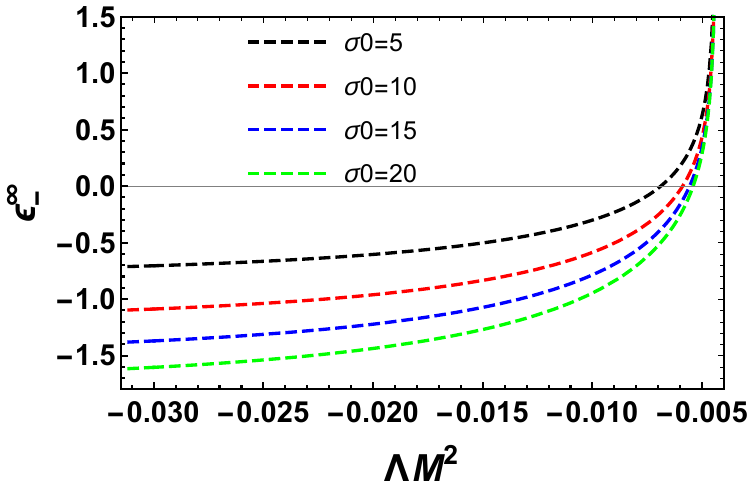}}
		\subfigure[\;$r/M=1.35, \xi=\pi/12$]{\label{4c}
			\includegraphics[width=4.8cm]{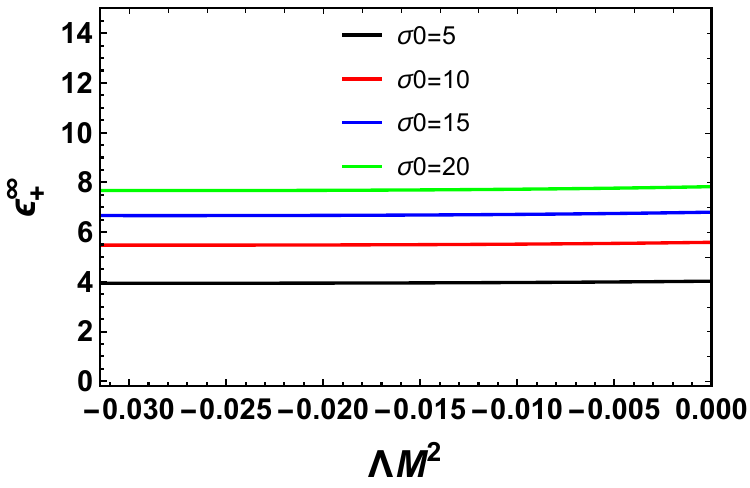}}
		\subfigure[\;$r/M=1.35, \xi=\pi/12$]{\label{4d}
			\includegraphics[width=4.8cm]{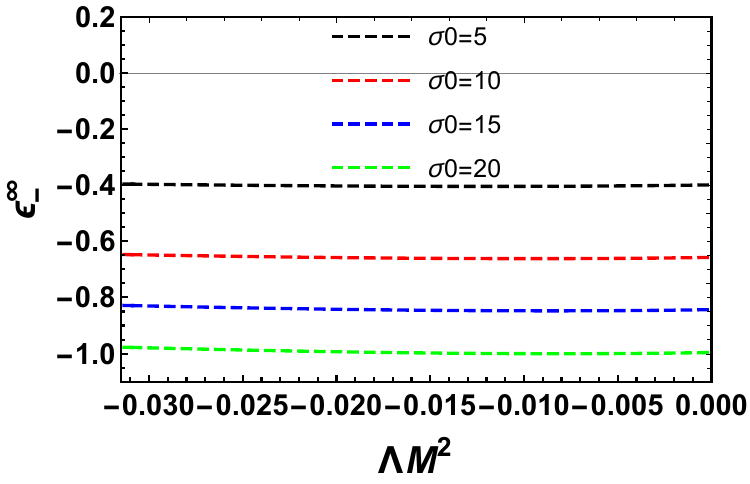}}
		\subfigure[\;$r/M=1.75, \xi=\pi/12$]{\label{4e}
			\includegraphics[width=4.8cm]{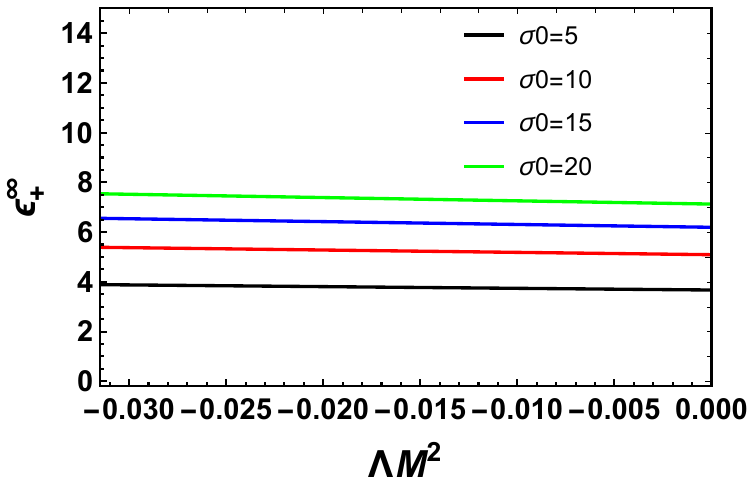}}
		\subfigure[\;$r/M=1.75, \xi=\pi/12$]{\label{4f}
			\includegraphics[width=4.8cm]{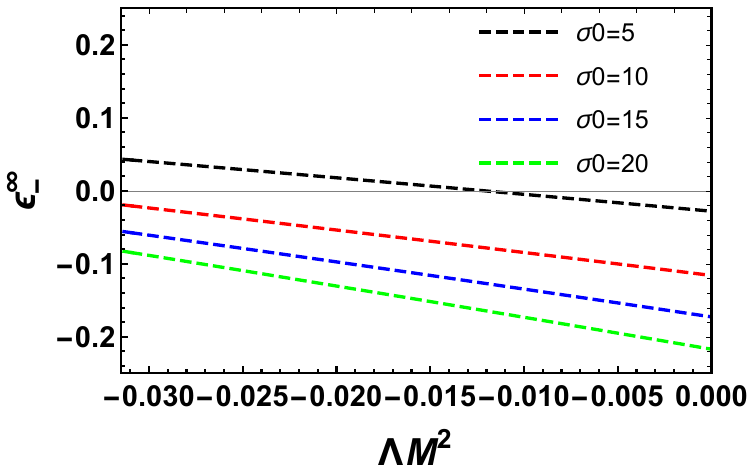}}
		\subfigure[\;$r/M=1.15, \xi=\pi/6$]{\label{4g}
			\includegraphics[width=4.8cm]{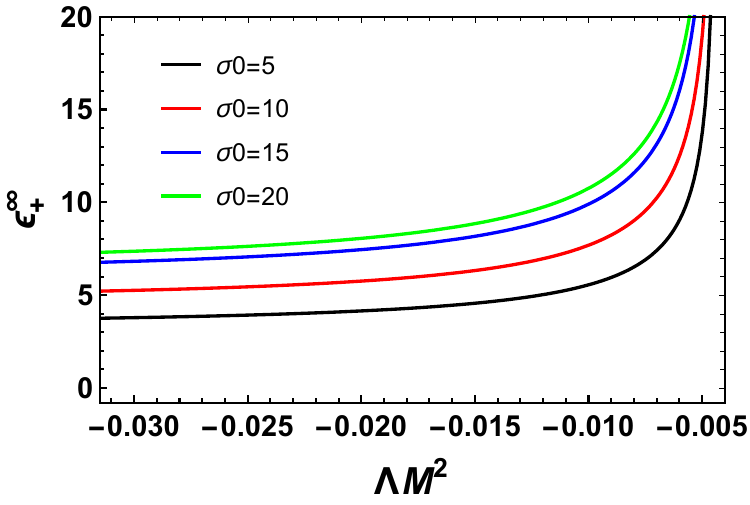}}
		\subfigure[\;$r/M=1.15, \xi=\pi/6$]{\label{4h}
			\includegraphics[width=4.8cm]{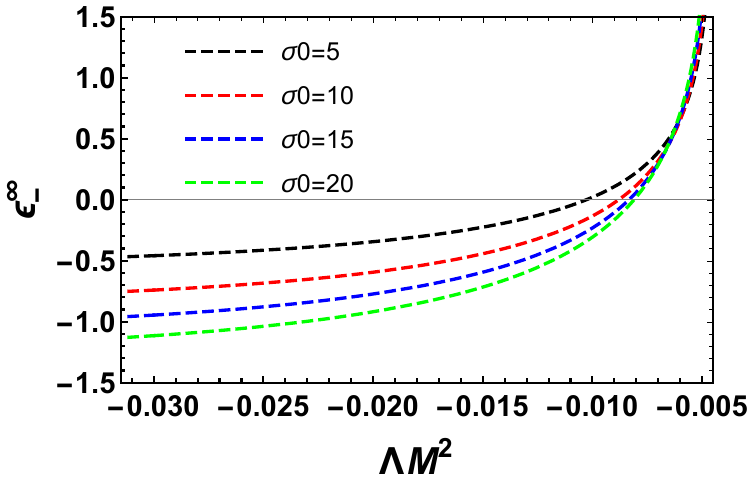}}
		\subfigure[\;$r/M=1.35, \xi=\pi/12$]{\label{4i}
			\includegraphics[width=4.8cm]{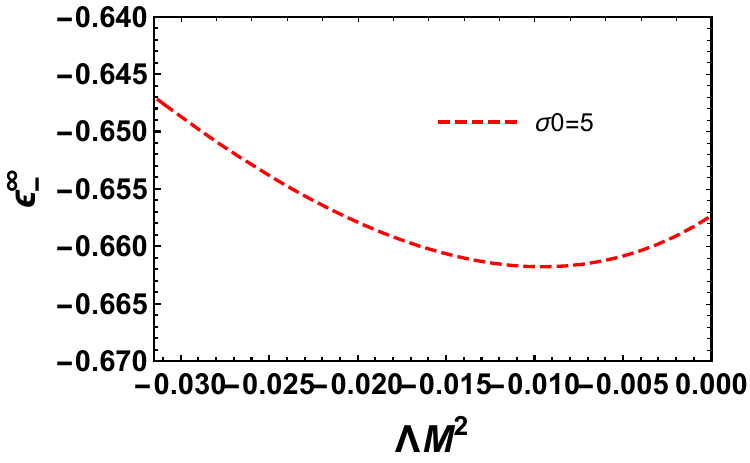}}
	\caption{The behaviors of $\epsilon^\infty_+$ (solid curves) and $\epsilon^\infty_-$ (dashed curves) as functions of the cosmological constant $\Lambda M^2$ for different values of the plasma magnetization of plasma $ \sigma_0$, the orientation angle $ \xi$, and the dominant reconnection radial location $r/M$, and with spin $a/M=0.99$ fixed. The plasma magnetization  $\sigma_0$ = 5, 10, 15, and 20 from bottom to top for solid curves and from top to bottom for dashed curves. (a) $r/M=1.15$, $\xi=\pi/12$. (b) $r/M=1.15$, $\xi=\pi/12$. (c) $r/M=1.35$, $\xi=\pi/12$. (d) $r/M=1.35$, $\xi=\pi/12$. (e) $r/M=1.75$, $\xi=\pi/12$. (f) $r/M=1.75$, $\xi=\pi/12$. (g) $r/M=1.15$, $\xi=\pi/6$. (h) $r/M=1.15$, $\xi=\pi/6$. (i) $r/M=1.35$, $\xi=\pi/12$. (i) shows the behavior of $\epsilon^\infty_{-}$ when $ \sigma_0$ is set to 5 in (d).}\label{fig4}
\end{figure}

It can be seen from Eq. \eqref{condition} that the energy-at-infinity per enthalpy of the plasma $\epsilon^\infty_{\pm}$ is a function of the black hole spin $a$, cosmological constant $\Lambda$, black hole mass $M$, the magnetization of plasma $ \sigma_0$, the orientation angle of plasma outflow velocity $ \xi$, and  the dominant reconnection radial location $r$. In order to investigate how these quantities affect energy extraction, we plot $\epsilon^\infty_+$ and $\epsilon^\infty_-$ in Fig.~\ref{fig3} and Fig.~\ref{fig4}, respectively.

In Fig. \ref{fig3}, we plot $\epsilon^\infty_{\pm}$ as functions of the spin parameter $a/M$ for varying values of the plasma magnetization $ \sigma_0$, the orientation angle $ \xi$, and the dominant reconnection radial location $r/M$, with the cosmological constant fixed at $\Lambda M^2=-0.01$. From Figs. \ref{3a}, \ref{3c}, \ref{3e}, and \ref{3g}, it reveals that $\epsilon^+_\infty$ remains strictly positive across all parameter variations. However, from Figs. \ref{3b}, \ref{3d}, \ref{3f}, and \ref{3h}, one can find that $\epsilon^-_\infty$ is negative when the black hole spin is beyond some certain values, which indicates that the conditions (\ref{condition}) are satisfied, and the energy extraction is possible. In particular, $\epsilon^\infty_-$ decreases with the spin $a/M$, which implies that more energy can be extracted for a larger spin.

As the plasma magnetization $\sigma_0$ increases from 5 to 20, with all other parameters held fixed, $\epsilon^\infty_-$  decreases, indicating that greater plasma magnetization leads to enhanced extraction of rotational energy from the black hole. A high degree of plasma magnetization implies the presence of a strong magnetic field configuration around the black hole, in which magnetic reconnection facilitates a more efficient conversion of stored magnetic energy into plasma kinetic energy, thereby enabling increased energy extraction from the black hole. A comparative analysis of Figs. \ref{3d} and \ref{3h}, conducted under fixed parameters, namely, a dominant reconnection radial location of $r/M=1.15$, identical black hole spin, and consistent plasma magnetization reveals that reducing the magnetic orientation angle $\xi$ from $\frac{\pi}{6}$ to $\frac{\pi}{12}$ results in lower $\epsilon^\infty_{-}$. This behavior signifies an increase in the amount of energy extracted from the black hole. These results show notable qualitative agreement with the conclusions drawn in Ref.~\cite{R18}, which investigates the influence of the Kerr black hole's and the plasma's physical parameters on energy extraction. Furthermore, in Figs. \ref{3b}, \ref{3d}, and \ref{3f}, we fix the magnetic orientation angle at $\xi=\frac{\pi}{12}$, while maintaining constant values for the black hole spin and plasma magnetization, and sequentially vary the dominant reconnection radial location to $r/M=1.1 $, $ r/M=1.4$, and $ r/M=1.6$. In these cases, the value of $\epsilon^+_\infty$ increases as $r/M$ grows.

We now turn to explore the dependence of the cosmological constant by plotting $\epsilon^\infty_{\pm}$ against $\Lambda M^2$ for $a/M=0.99$, and varying $\sigma_0$, $\xi$, $r/M$ in Fig. \ref{fig4}. In Figs. \ref{4a}, \ref{4c}, \ref{4e}, and \ref{4g}, $\epsilon^\infty_{+}$ consistently remains above 0. Therefore, our focus lies solely on $\epsilon^\infty_{-}$, depicted in Figs. \ref{4b}, \ref{4d}, \ref{4f}, and \ref{4h}. Notably, in Figs. \ref{4b} and \ref{4h}, with the dominant reconnection radial location set at $r/M=1.15$ and orientation angles $\xi=\frac{\pi}{6}$ and $\frac{\pi}{12}$ respectively, the results is consistent with the behaviors observed in Fig. \ref{fig3}, indicating that smaller values of $\xi$ lead to lower values of $\epsilon^\infty_{-}$. Similarly, we can draw a parallel to Fig. \ref{fig3} by concluding that an increase in plasma magnetization corresponds to a decrease in $\epsilon^\infty_{-}$. However, unlike that observed in Fig. \ref{fig3}, the comparison in subplots Fig. \ref{fig4}(b), 4(d), 4(f), and 4(i) reveals that $\epsilon^\infty_{-}$ does not exhibit a consistent monotonic increase or decrease with the rise in $\Lambda M^2$ at the dominant reconnection radial location $r/M$. For example, in Fig. \ref{4b}, with $r/M=1.15$ and $\xi=\frac{\pi}{12}$, $\epsilon^\infty_{-}$ demonstrates a continuous increase with $\Lambda M^2$. Contrastingly, in Figs. \ref{4d} and \ref{4i} (where Fig. \ref{4i} offers a clearer representation of the behavior of $\epsilon^\infty_{-}$ when $\sigma_0$ is set to 5 in Fig. \ref{4d}) with $r/M=1.35$ and $\xi=\frac{\pi}{12}$, a distinct behavior is observed where $\epsilon^\infty_{-}$ initially decreases, then increases with the increase in $\Lambda M^2$. In Fig. \ref{4f} with $r/M=1.75$ and $\xi=\frac{\pi}{12}$, $\epsilon^\infty_{-}$ decreases as $\Lambda M^2$ increases. This illustrates that the impact of the cosmological constant on energy extraction is influenced by the specific dominant reconnection radial location.

\section{Parameter spaces for energy extraction}
\label{paevmr}

In the preceding section, we calculate the hydrodynamic energy-at-infinity per enthalpy of the plasma, denoted as $\epsilon^\infty_{\pm}$, and the energy extraction condition outlined in \eqref{condition} through the magnetic reconnection mechanism. Obviously, the value of $\epsilon^\infty_{+}$ consistently remains positive. As a result, our primary focus concerning the energy extraction condition centers on ensuring $\epsilon^\infty_{-}<0$.

\begin{figure}
	\centering
	\subfigure[\;$\sigma_0=1$]{\label{5a}
		\includegraphics[width=5.1cm]{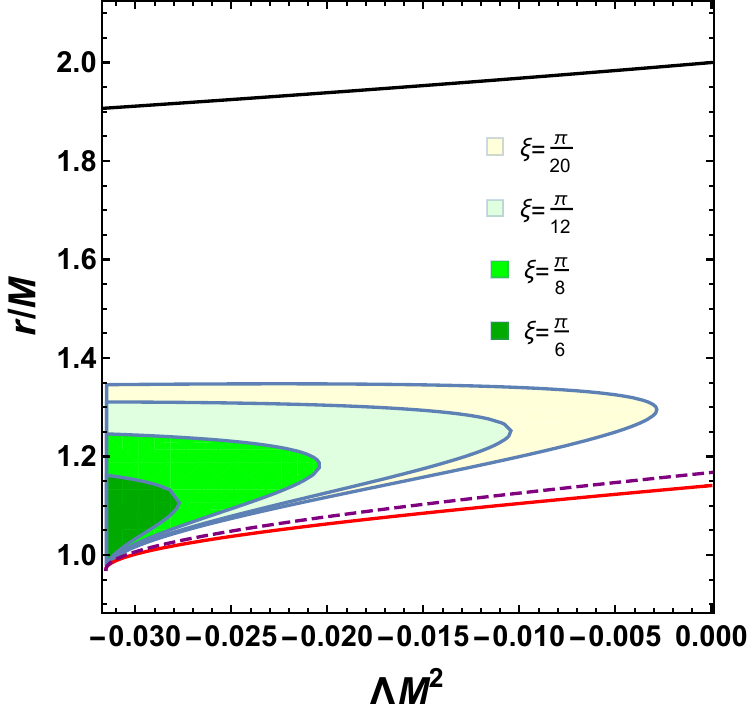}}
	\subfigure[\;$\xi=\frac{\pi}{12}$]{\label{5b}
		\includegraphics[width=5.1cm]{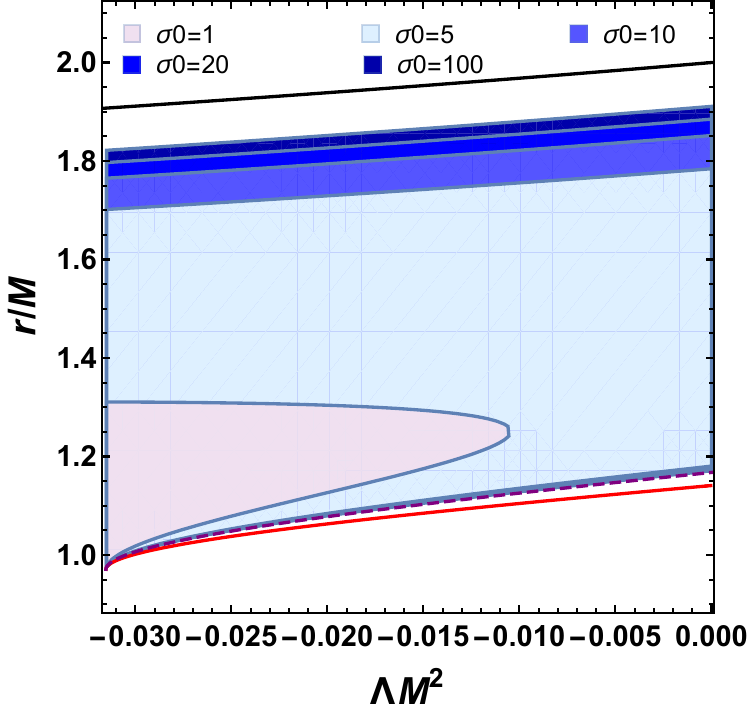}}
	\caption{Regions of parameter space $\Lambda M^2-r/M$ satisfying the energy extraction condition $\epsilon^\infty_{-}<0$ with $a/M=0.99$. Red solid curves, black solid curves, and purple dashed curves represent the radii of the outer event horizon $ r_+$, outer ergosphere boundary $ r_\text{E}$, and the circular corotating photon orbit $ r_{\text{LR}}$, respectively. (a) $\sigma_0=1$ with $\xi=\frac{\pi}{6}$, $\frac{\pi}{8}$, $\frac{\pi}{12}$, $\frac{\pi}{20}$. (b) $ \xi=\frac{\pi}{12}$ with $\sigma_0$=1, 5, 10, 20, 100.} \label{space}
\end{figure}

\begin{figure}
	\centering
	\subfigure[\;$\sigma_0=1$]{\label{6a}
		\includegraphics[width=5.1cm]{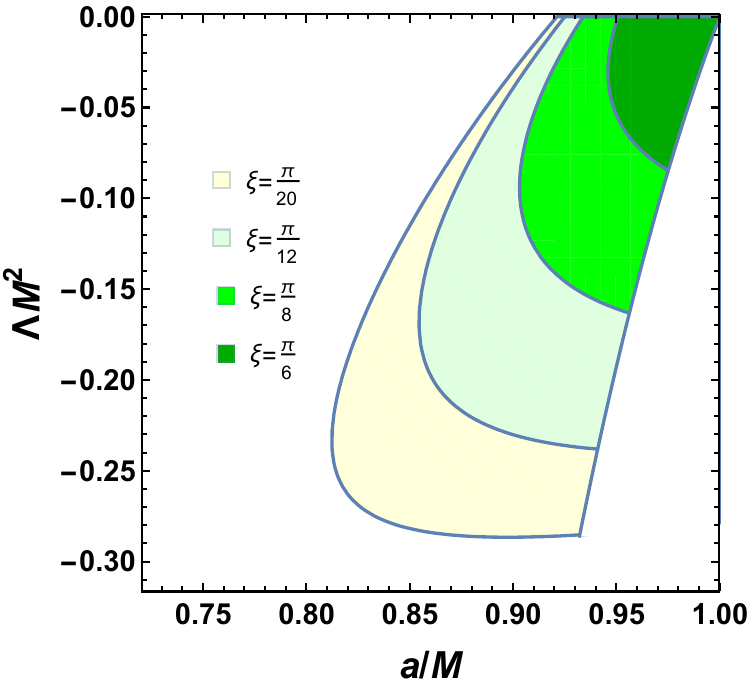}}
	\subfigure[\;$\xi=\frac{\pi}{12}$]{\label{6b}
		\includegraphics[width=5.0cm]{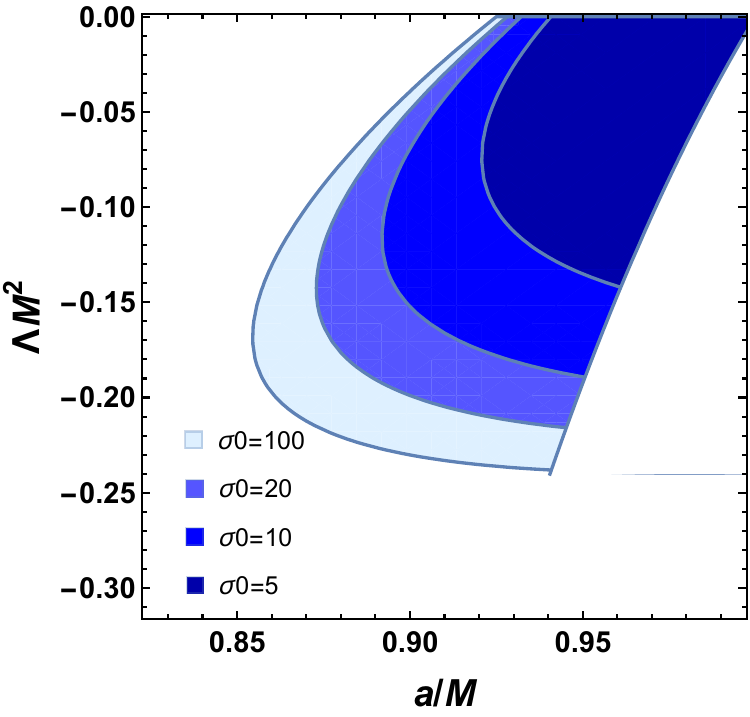}}
	\caption{Regions of parameter space $a/M-\Lambda M^2$ satisfying the energy extraction condition $\epsilon^\infty_{-}<0$ with $a/M=0.99$. The right boundaries of the shaded regions in both subplots correspond to the relationship curves between $a/M$ and $\Lambda M^2$ when the Kerr-AdS black hole is an extremal black hole. (a)  $\sigma_0=1$ with $\xi=\frac{\pi}{6}$, $\frac{\pi}{8}$, $\frac{\pi}{12}$, $\frac{\pi}{20}$. (b) $ \xi=\frac{\pi}{12}$ with $\sigma_0$=1, 5, 10, 20, 100.} \label{space2}
\end{figure}

This section delves into an examination of the parameter space. In Fig. \ref{space} and Fig. \ref{space2}, we present the regions within the parameter spaces of $\Lambda M^2-r/M$ and $a/M-\Lambda M^2$, respectively. These regions exhibit the configurations that satisfy the energy extraction condition $\epsilon^\infty_{-}<0$, offering a visual representation of how the cosmological constant influences the permissible range of physical parameters for energy extraction.

Firstly, in Fig. \ref{space}, we delineate the regions in the parameter space of $\Lambda M^2-r/M$ that satisfies the energy extraction condition $\epsilon^\infty_{\pm}<0$ for a rapidly rotating black hole with $a/M=0.99$. Moving to Fig. \ref{5a}, we explore the impact of the orientation angle $\xi$ under a constant plasma magnetization of $\sigma_0=1$. Shaded regions highlight negative values of $\epsilon^\infty_{\pm}$ for different $\xi$ values such as $\frac{\pi}{6}$, $\frac{\pi}{8}$, $\frac{\pi}{12}$, and $\frac{\pi}{20}$. Fig. \ref{5b} delves into the influence of various plasma magnetization levels ($\sigma_0=1$, 5, 10, 20, and 100) while maintaining a fixed orientation angle of $\xi=\frac{\pi}{12}$. This analysis reveals similar energy extraction parameter spaces across different magnetization values. Within the two subplots of Fig. \ref{space}, the red solid curves, black solid curves, and purple dashed curves correspond to the radii of the outer event horizon $r_+$, outer ergosphere boundary $r_\text{E}$, and the circular corotating photon orbit $r_{\text{LR}}$, as depicted in Fig. \ref{2a}. The shaded areas between $r_\text{E}$ and $r_{\text{LR}}$ signify the occurrence of magnetic reconnection within the ergosphere but beyond the photon orbit. Decreasing $\xi$ or increasing $\sigma_0$ expands these regions, broadening the permissible ranges of the parameter space in terms of $\Lambda M^2$ and $r/M$ for energy extraction. Importantly, the introduction of a negative cosmological constant systematically widens the acceptable range of the dominant reconnection radial location $r/M$ that complies with the energy extraction condition. This range consistently increases as the value of $\Lambda M^2$ decreases. Consequently, we infer that Kerr-AdS black holes facilitate energy extraction by plasmas across a broader radial spectrum for reconnection locations compared to their Kerr counterparts.

Subsequently, in Fig. \ref{space2}, we show the $a/M-\Lambda M^2$ parameter space to identify regions where $\epsilon^\infty_{\pm}<0$ while maintaining a fixed $r/M=1.5$. Moving to Fig. \ref{6a}, we present the outcomes for plasma magnetization at $\sigma_0=100$. Shaded regions denote negative values of $\epsilon^\infty_{\pm}$ corresponding to various orientation angles such as $\xi=\frac{\pi}{6}$, $\frac{\pi}{8}$, $\frac{\pi}{12}$, and $\frac{\pi}{20}$. Fig. \ref{6b}, with $\xi$ fixed at $\frac{\pi}{12}$, explores the effects of different plasma magnetization levels ($\sigma_0=5$, 10, 20, 100) on the parameter space. The boundaries of the shaded regions in both subplots align with the relationship curves between $a/M$ and $\Lambda M^2$ when the Kerr-AdS black hole represents an extremal black hole (as shown by the blue solid curve in Fig. \ref{fig1}). Observations reveal that a reduced orientation angle or an increased plasma magnetization augments the shaded regions, broadening the feasible ranges of $a/M$ and $\Lambda M^2$ for energy extraction. It is noteworthy that for fixed values of orientation angle $\xi$ or plasma magnetization $\sigma_0$, the viable range of $a/M$ for energy extraction initially expands and then contracts as $\Lambda M^2$ decreases from 0. This observation underscores that Kerr-AdS black holes empower plasma to extract energy even at significantly lower spin values (e.g., $a/M=0.86$), a feat unattainable by Kerr black holes ($\Lambda M^2=0$).

This discovery introduces an important opportunity for plasma to extract energy from black holes characterized by lower spin parameters, showcasing the unique capabilities of Kerr-AdS black holes in this energy extraction context.

\section{ Energy extraction efficiency and power}
\label{QQQ}

In the previous section, we calculate the hydrodynamic energy-at-infinity per enthalpy $\epsilon^\infty_{\pm}$ of the plasma, encompassing the vicinity of a Kerr-AdS black hole. Subsequently, we derived the conditions necessary for energy extraction, followed by an exploration of the parameter space that satisfies these energy extraction conditions, demonstrating the practicality of extracting energy from Kerr-AdS black holes. To further considering the plasma's capacity for extracting energy from a Kerr-AdS black hole, our investigation will now delve into evaluating the efficiency and power associated with this energy extraction process.

Now, we examine the efficiency of extracting energy from the Kerr-AdS black hole through the mechanism of magnetic reconnection. The efficiency of this energy extraction process can be defined by the following equation \cite{R18}
\begin{equation}
	\eta=\frac{\epsilon^\infty_+}{\epsilon^\infty_++\epsilon^\infty_-}.\label{conda}
\end{equation}
From \eqref{conda}, we can conclude that energy will be extracted from the Kerr-AdS black hole only when $\eta>1$.

\begin{figure}
	\centering
	\subfigure[\;$\Lambda M^2=-0.01$ ]{\label{7a}
		\includegraphics[width=5.85cm]{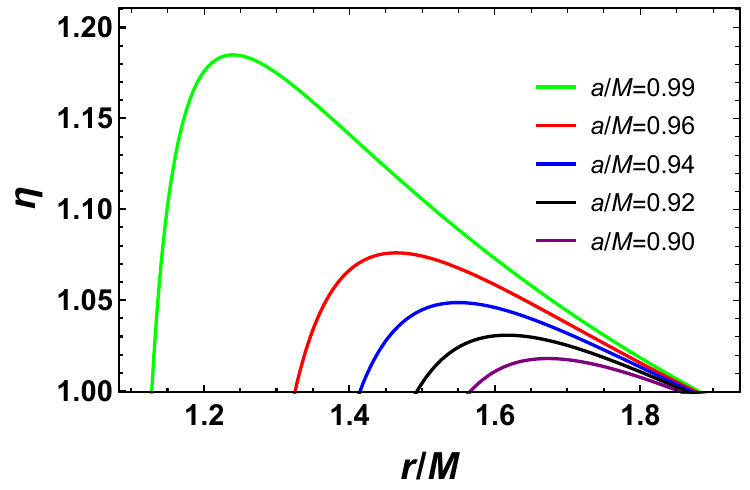}}
	\subfigure[\;$a/M=0.99$]{\label{7b}
		\includegraphics[width=5.75cm]{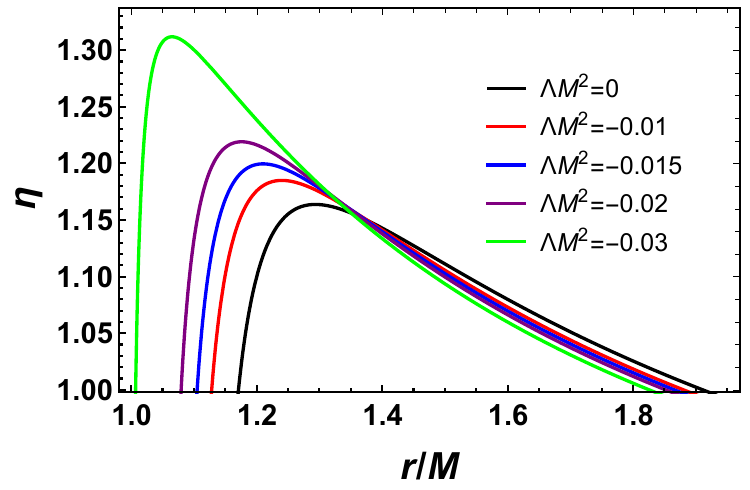}}
	\caption{(a) The behaviors of $\eta$ as a function of the dominant reconnection radial location $r/M$ by taking $\sigma_0=100$,  $\xi=\frac{\pi}{12}$ and $\Lambda M^2=-0.01$ with different values of the spin $a/M$. The spin $a/M$ = 0.99, 0.96, 0.94, 0.92 and 0.90 from top to bottom. (b) The behaviors of $\eta$ as a function of the dominant reconnection radial location $r/M$ by taking $\sigma_0=100$,  $\xi=\frac{\pi}{12}$ and $a/M=0.99$ with different values of the cosmological constant $\Lambda M^2$. The cosmological constant $\Lambda M^2$ = -0.03, -0.02, -0.015, -0.01 and 0 from top to bottom. } \label{xiaolv1}
\end{figure}

Building on the insights from Ref. \cite{R18}, we investigate the efficiency $\eta$ of energy extraction as a function of the dominant reconnection radial position $r/M$ for Kerr-AdS black holes, maintaining a constant plasma magnetization of $\sigma_0=100$ and reconnection angle $\xi=\frac{\pi}{12}$. In Fig. \ref{7a}, with the cosmological constant fixed at $\Lambda M^2=-0.01$, the efficiency $\eta$ displays non-monotonic behavior across all tested spin values $a/M$. It exhibits an initial ascent followed by a descent as $r/M$ increases, with peak values occurring in proximity to the circular corotating photon orbit. Notably, higher values of $a/M$ correspond to larger peak efficiencies $\eta$, underlining the enhanced energy extraction potential with increasing black hole spin.

Examining Fig. \ref{7b} with $a/M=0.99$, a similar behavior is observed where the efficiency $\eta$ initially rises and then decreases as $r/M$ varies across different cosmological constant values $\Lambda M^2$. The peak efficiency for $\eta$ is consistently obtained when the dominant reconnection radial locations are closer to the circular corotating photon orbit. Furthermore, lower values of $a/M$ correspond to higher peak efficiencies $\eta$. Remarkably, the Kerr black hole scenario ($\Lambda M^2=0$) achieves the smallest peak efficiency $\eta$. This comparison underscores the significantly enhanced energy extraction capability of plasma in Kerr-AdS black holes compared to Kerr black holes, particularly when the reconnection occurs in the vicinity of critical photon orbits. This enhanced energy extraction strength further accentuates as the cosmological constant $\Lambda M^2$ decreases.

\begin{figure}
	\centering
	\subfigure[\;$\sigma_0=100$,  $\xi=\frac{\pi}{12}$, $a/M=0.99$]{\label{8a}
		\includegraphics[width=5.8cm]{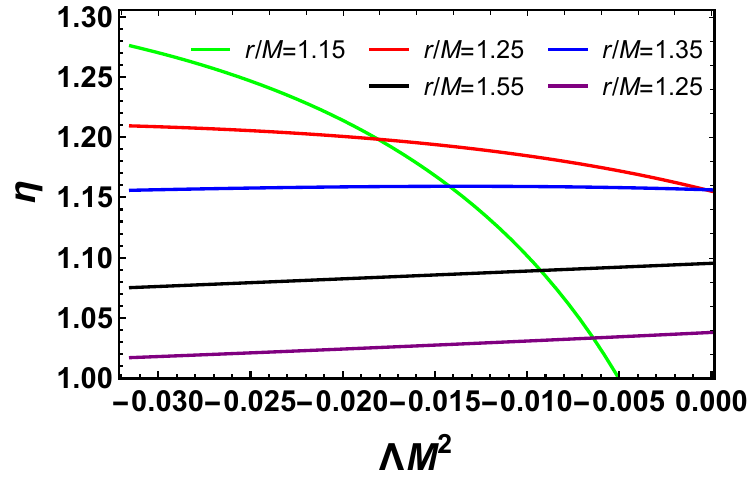}}
	\subfigure[\;$\sigma_0=100$,  $\xi=\frac{\pi}{12}$, $a/M=0.99$]{\label{8b}
		\includegraphics[width=5.8cm]{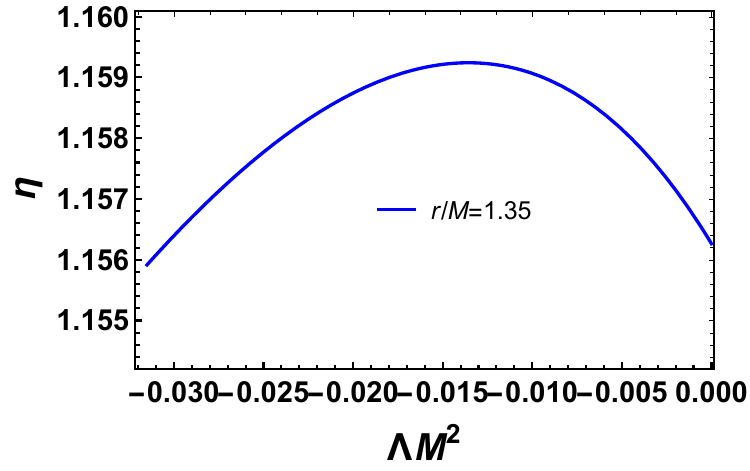}}
	\caption{The behavior of efficiency $\eta$ as a function of the cosmological constant $\Lambda M^2$ by taking $\sigma_0=100$,  $\xi=\frac{\pi}{12}$ and $a/M=0.99$ with different dominant reconnection radial location $r/M$. (b) shows the behavior of $\eta$ when $ r/M$ is set to 1.35 in (a).} \label{xiaolv2}
\end{figure}

Subsequently, in Fig. \ref{xiaolv2}, we set the spin at $a/M = 0.99$, plasma magnetization at $\sigma_0=100$, and the orientation angle at $\xi=\frac{\pi}{12}$ to examine the efficiency $\eta$ as a function of the cosmological constant $\Lambda M^2$ across various dominant reconnection radial locations $r/M$. The efficiency demonstrates distinct behaviors with the specific radial position of dominant reconnection. At smaller radii, like $r/M=1.25$, $\eta$ exhibits a continuous decrease as $\Lambda M^2$. When the radial position slightly increases, for example, at $r/M=1.35$, $\eta$ initially rises and then falls with $\Lambda M^2$, as depicted in Fig. \ref{8b}. In contrast, at larger radial positions such as $r/M=1.75$, $\eta$ consistently rises with $\Lambda M^2$. Nonetheless, across all scenarios, a lower cosmological constant correlates with higher efficiency $\eta$, indicating a more robust energy extraction capability. These results emphasize the significance of the dominant reconnection radial position in modulating the efficacy of energy extraction, with lower radial positions favoring heightened energy extraction. However, when $r/M$ is extremely small, such as at 1.15, the value of $\eta$ precipitously drops below one as the cosmological constant increases.

\begin{figure}[htb]
	\centering
	\subfigure[\;$\sigma_0=10000$, $\Lambda M^2=0$]{\label{9a}
		\includegraphics[width=4.8cm]{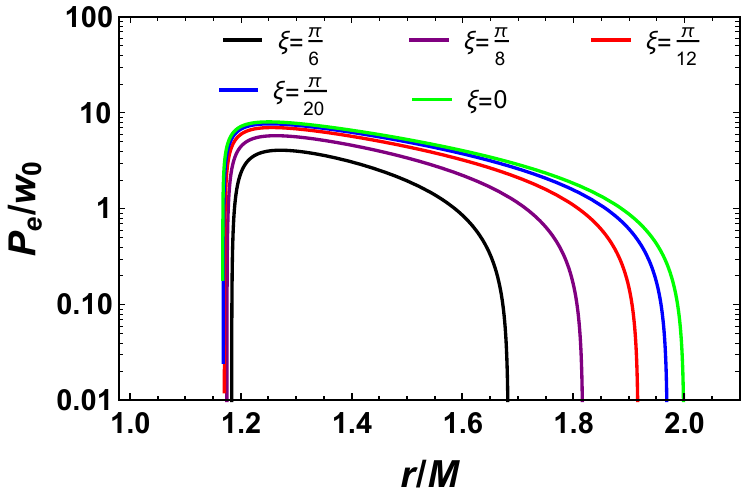}}
	\subfigure[\;$\sigma_0=10000$, $\Lambda M^2=-0.03$]{\label{9b}
		\includegraphics[width=4.8cm]{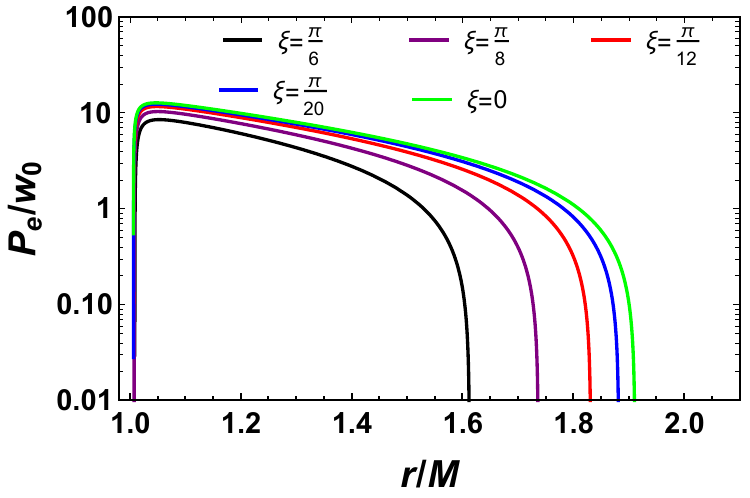}} \\
	\subfigure[\;$\xi=\frac{\pi}{12}$, $\Lambda M^2=0$]{\label{9c}
		\includegraphics[width=4.8cm]{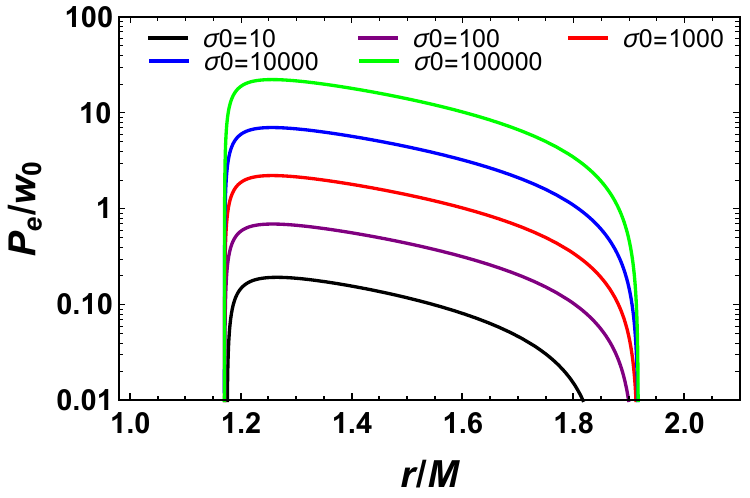}}
	\subfigure[\;$\xi=\frac{\pi}{12}$, $\Lambda M^2=-0.03$]{\label{9d}
		\includegraphics[width=4.8cm]{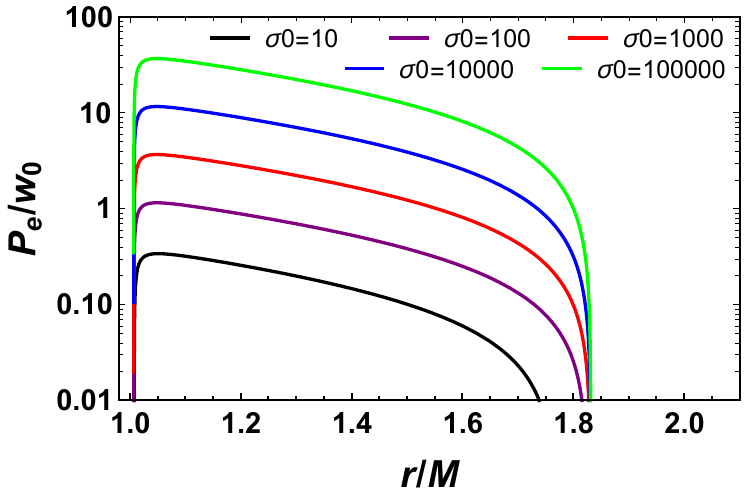}}
	\caption{Log-plot of the power $P_{\text{extr}}/\omega_0$ as a function of the dominant reconnection radial location $r/M$ with $a/M=0.99$ and $\Lambda M^2=-0.01$. (a), (b) The plasma magnetization $\sigma_0=10$, and the orientation angle $\xi= \frac{\pi}{6}$, $\frac{\pi}{8}$, $\frac{\pi}{12}$, $\frac{\pi}{20}$, 0 from bottom to top.  (c), (d) the orientation angle $\xi=\frac{\pi}{12}$, and the plasma magnetization $\sigma_0=10$, 100, 1000, 10000, 100000 from bottom to top.} \label{gonglv1}
\end{figure}

Another crucial aspect in evaluateing the potential for energy extraction from Kerr-AdS black holes via magnetic reconnection is the power of energy extraction. The power $P_{\text{extr}}$ denotes the rate at which the plasma escapes the black hole to extract energy from it and can be defined as \cite{R18}
\begin{equation}
	P_{\text{extr}}=-\epsilon^\infty_-\omega_0A_{\text{in}}U_{\text{in}},\label{shizi}
\end{equation}
where $A_{\text{in}}$ denotes the cross-sectional area of the inflowing plasma, which for rapidly spinning black holes is approximately $A_{\text{in}}\sim(r_\text{E}^2-r_{\text{LR}}^2)$. The plasma inflow four-velocity is represented by $U_{\text{in}}$, typically $U_{\text{in}}=\mathcal{O}(10^{-1})$ for the collisionless regime and $U_{\text{in}}=\mathcal{O}(10^{-2})$ for the collisional regime, as discussed in \cite{R77}, with $U_{\text{in}}$ set at 0.1 for this analysis. Consequently, the power per unit enthalpy can be furhter expressed as
\begin{equation}
	P_{\text{extr}}/\omega_0=-0.1\epsilon^\infty_-(r_\text{E}^2-r_{\text{LR}}^2).
\end{equation}

We consider a black hole spin of $a/M=0.99$ and illustrate the power $P_{\text{extr}}/\omega_0$ as a function of the dominant reconnection radial location $r/M$ in Fig. \ref{gonglv1}. Figs. \ref{9a} and \ref{9b} display the behavior of $P_{\text{extr}}/\omega_0$ with increasing $r/M$ for various orientation angles $\xi$ under cosmological constant values of $\Lambda M^2=0$ and $\Lambda M^2=-0.03$, respectively. Additionally, Figs. \ref{9c} and \ref{9d} showcase the behavior of $P_{\text{extr}}/\omega_0$ with increasing $r/M$ for different plasma magnetization $\sigma_0$ under cosmological constant values of $\Lambda M^2=0$ and $\Lambda M^2=-0.03$, respectively. Note that $\Lambda M^2=0$ denotes the scenario of a Kerr black hole. From these subplots, it is evident that the power $P_{\text{extr}}/\omega_0$ peaks at an $r/M$ value in close proximity to the dominant reconnection radial location near the circular corotating photon orbit, following which it decreases with further increments in $r/M$. A smaller orientation angle $\xi$ or a higher plasma magnetization $\sigma_0$ correlates with increased energy extraction power $P_{\text{extr}}/\omega_0$. Comparative analysis between plots with $\Lambda M^2=0$ and $\Lambda M^2=-0.03$ reveals that a reduction in the cosmological constant $\Lambda M^2$ reduces the radius of the co-rotating photon circular orbit, shifting the peak towards smaller $r/M$ values while intensifying the peak of $P_{\text{extr}}/\omega_0$. This observation strongly suggests that the presence of a negative cosmological constant enhances the plasma's capability to extract energy from the black hole.

\begin{figure}
	\centering
	\subfigure[\;$\sigma_0=10000$, $\xi=\frac{\pi}{12}$, $a/M=0.99$]{\label{10a}
		\includegraphics[width=5.8cm]{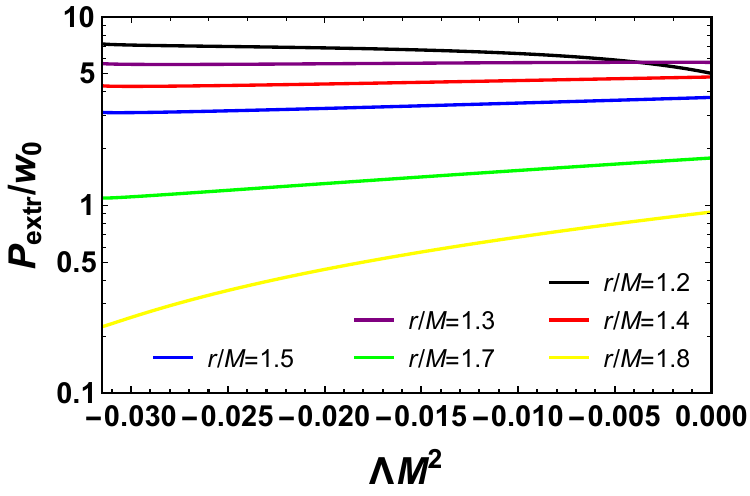}}
	\subfigure[\;$\sigma_0=10000$, $\xi=\frac{\pi}{12}$, $a/M=0.99$]{\label{10b}
		\includegraphics[width=5.8cm]{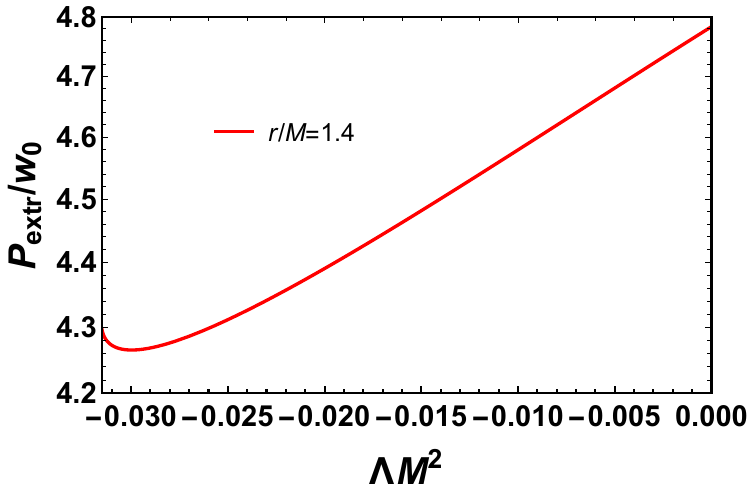}}
	\caption{(a) Lot-plot of power $P_{\text{extr}}/\omega_0$ as a function of cosmological constant $\Lambda M^2$ with different dominant reconnection radial location by taking $a/M=0.99$, $\sigma_0=10000$ and $\xi=\frac{\pi}{12}$. (b) shows the behavior of $P_{\text{extr}}/\omega_0$ when $r/M=1.4$ in (a).} \label{gonglv3}
\end{figure}

To further illustrate the impact of the cosmological constant on the power dynamics, we present the power $P_{\text{extr}}/\omega_0$ as a function of the cosmological constant $\Lambda M^2$ across different dominant reconnection radial locations $r/M$. This analysis is conducted with $a/M=0.99$, $\sigma_0=10000$, and $\xi=\frac{\pi}{12}$, as depicted in Fig. \ref{gonglv3}. Fig. \ref{10b} provides a detailed breakdown of the power behavior in relation to $\Lambda M^2$ when $r/M$ is set to 1.4, as shown in Fig. \ref{10a}. When $r/M$ is relatively small, e.g., $r/M = 1.2$, the power $P_{\text{extr}}/\omega_0$ exhibits a consistent decrease with increasing $\Lambda M^2$. At $r/M = 1.4$, Fig. \ref{10b} illustrates that as $\Lambda M^2$ increases, $P_{\text{extr}}/\omega_0$ initially decreases before transiting to an increase. Considering larger values of $r/M$, exemplified by $r/M=1.8$, $P_{\text{extr}}/\omega_0$ demonstrates a continuous increase alongside the increase in $\Lambda M^2$. In general, when the dominant reconnection radial location $r/M$ is smaller, the power $P_{\text{extr}}/\omega_0$ tends to be higher, with lower values of $\Lambda M^2$ corresponding to increased $P_{\text{extr}}/\omega_0$. This observation underscores the intricate relationship between the dominant reconnection radial position, the cosmological constant, and the power dynamics of energy extraction from Kerr-AdS black holes.

Finally, for a comprehensive comparison of the energy extraction efficiency between plasma in Kerr-AdS black holes and Kerr black holes via the magnetic reconnection mechanism, we introduce the energy extraction power ratio $\kappa$ defined as: \cite{R18}
\begin{equation}
	\kappa=\frac{P_{\text{extr}}^{\text{Kerr-AdS}}}{P_{\text{extr}}^{\text{Kerr}}}.
\end{equation}
For $P_{\text{extr}}^{\text{Kerr}}$, we only need to substitute $\Lambda M^2=0$ into \eqref{shizi}.

\begin{figure}
	\center{\includegraphics[width=5.8cm]{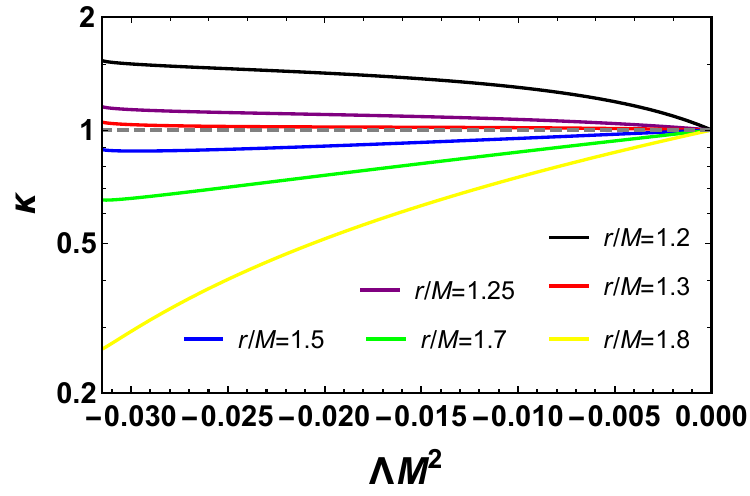}}
	\caption{The behavior of power ratio $\kappa$ as a function of cosmological constant $\Lambda M^2$ with different dominant reconnection radial location by taking $a/M=0.99$, $\sigma_0=10000$, and $\xi=\frac{\pi}{12}$.} \label{gonglv4}
\end{figure}

In Fig. \ref{gonglv4}, we fix the spin at $a/M = 0.99$, $\sigma_0=10000$ and $\xi= \frac{\pi}{12}$, and then illustrate the power ratio $\kappa$ as a function of the cosmological constant $\Lambda M^2$ across various dominant reconnection radial locations $r/M$. The behavior reveals that for smaller dominant reconnection radial locations, such as $r/M<1.5$, $\kappa$ surpasses one, with a smaller cosmological constant $\Lambda M^2$ resulting in a higher $\kappa$. This observation suggests that energy extraction from a Kerr-AdS black hole provides more efficient than that from a Kerr black hole, with the presence of the cosmological constant $\Lambda M^2$ amplifying the efficacy of magnetic reconnection for energy extraction. However, when $r/M$ is relatively large, as exemplified by the value 1.7, $\kappa<1$. In such instances, the Kerr-AdS black hole does not offer a notable advantage in energy extraction compared to a Kerr black hole. In conclusion, the dominant reconnection radial location $r/M$ significantly impacts the energy extraction from Kerr-AdS black holes through the magnetic reconnection mechanism. Notably, plasma demonstrates a substantial edge in energy extraction at smaller $r/M$ positions within Kerr-AdS black holes in contrast to Kerr black holes.

\begin{figure}
	\center{\includegraphics[width=5.8cm]{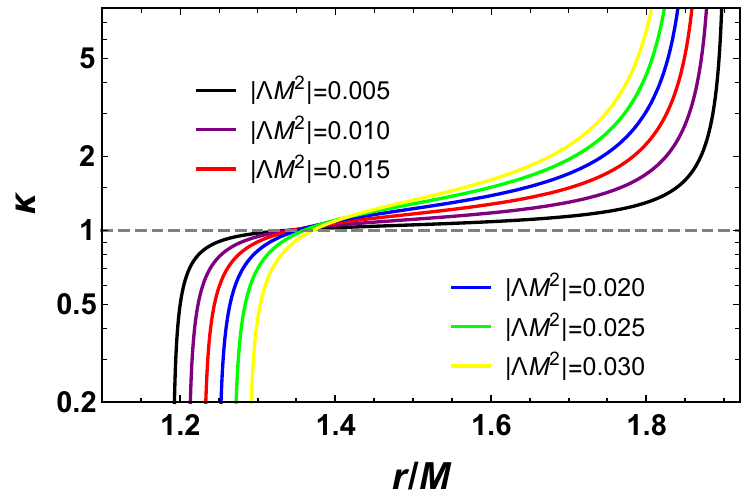}}
	\caption{The behavior of power ratio $\kappa$ as a function of the dominant reconnection radial location $r/M$ with values of different cosmological constant by taking $a/M=0.99$, $\sigma_0=10000$, and $\xi=\frac{\pi}{12}$.} \label{gonglv5}
\end{figure}

On the other hand, in order to further highlight the unique role of the cosmological constant on energy extraction by magnetic reconnection, we calculate the following energy extraction power ratio to compare the energy extraction power of Kerr-dS black holes with a positive cosmological constant and Kerr-AdS black holes with a negative cosmological constant,
\begin{equation}
 \kappa=\frac{P_{\text{extr}}^{\text{Kerr-dS}}}{P_{\text{extr}}^{\text{Kerr-AdS}}},
\end{equation}
where $P_{\text{extr}}^{\text{Kerr-dS}}$ has been discussed in detail in Ref. \cite{R33}.

By fixing $a/M = 0.99$, $\sigma_0=10000$ and $\xi= \frac{\pi}{12}$ in Fig. \ref{gonglv5}, we observe the power ratio $\kappa$ as a function of the dominant reconnection radial location $r/M$. Curves of different colors represent Kerr-dS and Kerr-AdS black holes with different values of the cosmological constant. For example, $|\Lambda M^2|=0.03$ denotes $\Lambda M^2=0.03$ for Kerr-dS black holes and $\Lambda M^2=-0.03$ for Kerr-AdS black holes. From Fig. \ref{gonglv5}, we can see that as the dominant reconnection radial location increases, $\kappa$ shows an upward trend. This indicates that during the increase of the dominant reconnection radial location, the advantage of energy extraction in Kerr-dS black holes gradually improves compared to that in Kerr-AdS black holes. However, when the dominant reconnection radial location is relatively small, such as $r/M<1.35$, $\kappa$ is consistently less than 1. This suggests that in Kerr-AdS black holes, the energy extraction power is greater, making energy extraction more favorable than in Kerr-dS black holes. Moreover, at smaller $r/M$, when the cosmological constant $\Lambda M^2$ for Kerr-AdS is smaller, for example $\Lambda M^2=-0.03$, $\kappa$ becomes even smaller. In this case, the advantage of energy extraction in Kerr-AdS black holes is further enhanced compared to Kerr-dS black holes. In contrast, at larger dominant radial positions, such as $r/M>1.7$, $\kappa$ is significantly greater than 1. Here, the advantage of energy extraction in Kerr-dS black holes is markedly stronger than in Kerr-AdS black holes, and for Kerr-dS black holes, a larger cosmological constant $\Lambda M^2$ corresponds to a larger $\kappa$.

Above, we calculated the ratio of energy extraction power, comparing the power of energy extraction in Kerr-AdS black holes with that in Kerr black holes and Kerr-dS black holes, respectively. We conclude that when the dominant reconnection radial location of magnetic reconnection relatively small and the negative cosmological constant $\Lambda M^2$ is smaller, energy extraction in Kerr-AdS black holes shows a significant advantage compared to Kerr and Kerr-dS black holes.

\section{Conclusions and discussions}
\label{Conclusion}

As predicted by general relativity, spinning black holes possess considerable energy. Since Penrose introduced the Penrose process for extracting rotational energy from spinning black holes, various energy extraction mechanisms have been explored to elucidate high-energy astrophysical phenomena. Comisso and Asenjo conducted an in-depth study on extracting energy from Kerr black holes using magnetic reconnection. This study extends the investigation to Kerr-AdS black holes with negative cosmological constants, examining the viability of energy extraction through magnetic reconnection in this context.

Given that energy extraction occurs within a black hole's ergosphere, we initially analyzed the spacetime structure and characteristic radii of Kerr-AdS black holes. By studying the event horizon, we delineated the parameter space $\Lambda M^2-a/M$ allowing for Kerr-AdS black hole existence. Notably, as the spin decreases, the minimum negative cosmological constant value also decreases, approaching negative infinity with a Kerr-AdS black hole spin nearing 0. Study on the outer ergosphere boundary and circular corotating photon orbit reveals how these characteristic radii vary with $\Lambda M^2$ or $a/M$. The radii values increase with cosmological constant under fixed spin conditions, and decrease with spin under fixed cosmological constant.

Subsequently, we calculated the hydrodynamic energy-at-infinity per plasma enthalpy $\epsilon^\infty_{\pm}$ around Kerr-AdS black holes, deriving conditions for energy extraction. Plotting $\epsilon^\infty_{+}$ and $\epsilon^\infty_{-}$ as functions of spin $a/M$, cosmological constant $\Lambda M^2$, magnetization $\sigma_0$, orientation angle $\xi$, and dominant reconnection radial location $r/M$, emphasizes the dependence of Kerr-AdS black hole energy extraction via magnetic reconnection on $\epsilon^\infty_{-}$. Notably, greater spin $a/M$, higher plasma magnetization $\sigma_0$, and smaller azimuthal angles $\xi$ enable more efficient energy extraction, akin to observations in Kerr black holes. The impact of cosmological constant $\Lambda M^2$ on energy extraction is influenced by dominant reconnection radial location $r/M$. Notably, at smaller $r/M$, energy extraction from Kerr-AdS black holes surpasses Kerr black holes, with extractable energy increasing as $\Lambda M^2$ decreases.

Then we delineated parameter space regions $\Lambda M^2-r/M$ and $a/M-\Lambda M^2$ satisfying energy extraction conditions. Higher plasma magnetization $\sigma_0$ and lower orientation angle $\xi$ widen the permissible ranges of spin $a/M$, cosmological constant $\Lambda M^2$, and dominant reconnection radial location $r/M$ for energy extraction from Kerr-AdS black holes. Negative cosmological constants expand the feasible $a/M$ and $r/M$ ranges, enabling energy extraction from low-spin black holes.

To further evaluate plasma's energy extraction capability from Kerr-AdS black holes through magnetic reconnection, we calculated the efficiency $\eta$ and power $P_{\text{extr}}/\omega_0$. Enhanced plasma magnetization $\sigma_0$ and reduced orientation angle $\xi$ lead to higher efficiency and power. Notably, energy extraction power and efficiency peak at smaller $r/M$, with efficiencies not consistently increasing or decreasing with $\Lambda M^2$ changes. By comparing energy extraction capabilities of Kerr-AdS with that of Kerr black holes and Kerr-dS black holes via magnetic reconnection using power ratio $\kappa$, we found that at smaller dominant reconnection radial locations, compared with Kerr black holes and Kerr-dS black holes, Kerr-AdS black holes exhibit a more superior energy extraction advantage. Moreover, as the negative cosmological constant of the Kerr-AdS black hole decreases, the energy extraction capability further enhances.

In summary, this study demonstrates the feasibility of extracting energy from Kerr-AdS black holes and underscores the significant impact of negative cosmological constants on black hole energy extraction via magnetic reconnection. The presence of negative cosmological constants expands the permissible dominant reconnection radial locations for meeting energy extraction conditions, facilitating energy extraction from low-spin black holes. The influence of negative cosmological constants on energy extraction is modulated by dominant reconnection radial location, with more energy extractable at smaller $r/M$, particularly as $\Lambda M^2$ decreases. Energy extraction from Kerr-AdS black holes proves more advantageous than from Kerr black holes and kerr-dS black holes via magnetic reconnection.

Future study may explore energy extraction from Kerr-AdS black holes through the collisional Penrose process, BSW acceleration mechanism, and the Blandford-Znajek mechanism, by comparing these with magnetic reconnection-based energy extraction. Additionally, investigations into energy extraction from other rotating black hole types, such as higher-dimensional rotating black holes \cite{R78}, could offer further insights into energy extraction mechanisms.

\section*{Acknowledgments}
This work was supported by the National Natural Science Foundation of China (Grants No. 12475055, and No. 12247101), the Basic Research Foundation of Central Universities (Grant No. lzujbky-2024-jdzx06), the Natural Science Foundation of Gansu Province (Grant No. 22JR5RA389), and the `111 Center' under Grant No. B20063.

\end{document}